\documentclass[prb,twocolumn,superscriptaddress,amsart,longbibliography]{revtex4}

\usepackage{graphicx, tikz-cd}
\usepackage{multirow}
\usepackage{bm}
\usepackage{times}
\usepackage{amsmath,bm,amsfonts}
\usepackage{dcolumn}
\usepackage{graphicx}
\usepackage{latexsym}
\usepackage{bbm,amsthm,amssymb,verbatim,tikz,physics}
\usepackage{ulem} 
\usepackage{braket}
\usepackage{bbold}

\graphicspath{{Figures/}}
\usepackage{bbm}
\usepackage{mathtools}
\usepackage{leftindex}
\usepackage{tikz}

\usepackage{natbib}

\usepackage{xcolor, soul}

\makeatletter
\newcommand*{\rom}[1]{\expandafter\@slowromancap\romannumeral #1@}
\makeatother
\newcommand\encircle[1]{%
  \tikz[baseline=(X.base)] 
    \node (X) [draw, shape=circle, inner sep=0] {\strut #1};}

\setcitestyle{square,numbers}

\usepackage{hyperref}
\hypersetup{
    colorlinks=true,
    linkcolor=blue,
    filecolor=blue,      
    urlcolor=blue,
    citecolor=blue,
}

\begin{document}

\preprint{APS/123-QED}

\title{Weak localization in magnetic Euler bands}

\author{Doh Young \surname{Kim}}
\affiliation{Department of Physics and Astronomy, Seoul National University, Seoul 08826, Korea}
\affiliation{Center for Theoretical Physics (CTP), Seoul National University, Seoul 08826, Korea}
\affiliation{Institute of Applied Physics, Seoul National University, Seoul 08826, Korea}

\author{Seung Hun \surname{Lee}}
\affiliation{Department of Physics and Astronomy, Seoul National University, Seoul 08826, Korea}
\affiliation{Center for Theoretical Physics (CTP), Seoul National University, Seoul 08826, Korea}
\affiliation{Institute of Applied Physics, Seoul National University, Seoul 08826, Korea}

\author{Akira \surname{Furusaki}}
\email{furusaki@riken.jp}
\affiliation{RIKEN Center for Emergent Matter Science, Wako, Saitama, 351-0198, Japan}

\author{Bohm-Jung \surname{Yang}}
\email{bjyang@snu.ac.kr}
\affiliation{Department of Physics and Astronomy, Seoul National University, Seoul 08826, Korea}
\affiliation{Center for Theoretical Physics (CTP), Seoul National University, Seoul 08826, Korea}
\affiliation{Institute of Applied Physics, Seoul National University, Seoul 08826, Korea}

\date{\today}

\begin{abstract}
We study the quantum correction to the conductivity due to disorder in two-dimensional fragile topological
bands with nonzero Euler class. Contrary to graphene where two Dirac nodes have opposite vorticities,
two bands with a unit Euler number possess two Dirac points with the same vorticity, which may affect the
Anderson localization.
Most notably, we report an anomalous localization behavior in spinful magnetic Euler bands based on symmetry analysis and diagrammatic calculations. Despite the presence of spin-orbit coupling and an in-plane magnetization that explicitly breaks \textit{physical} time-reversal symmetry, the system exhibits weak localization behavior characteristic of the orthogonal symmetry class. 
We demonstrate that this counter-intuitive phenomenon originates from an emergent \textit{effective} time-reversal symmetry composed of crystalline and spacetime inversion symmetries, allowing it to supersede the standard localization behavior. 
Our findings reveal that the \textit{effective} crystalline symmetries can fundamentally alter the universality class of disordered systems, rendering the localization behavior independent of the specific vorticity configuration of Dirac nodes.
\end{abstract}

\maketitle

\begin{figure}[!b]
\centering
\includegraphics[width=0.5\textwidth]{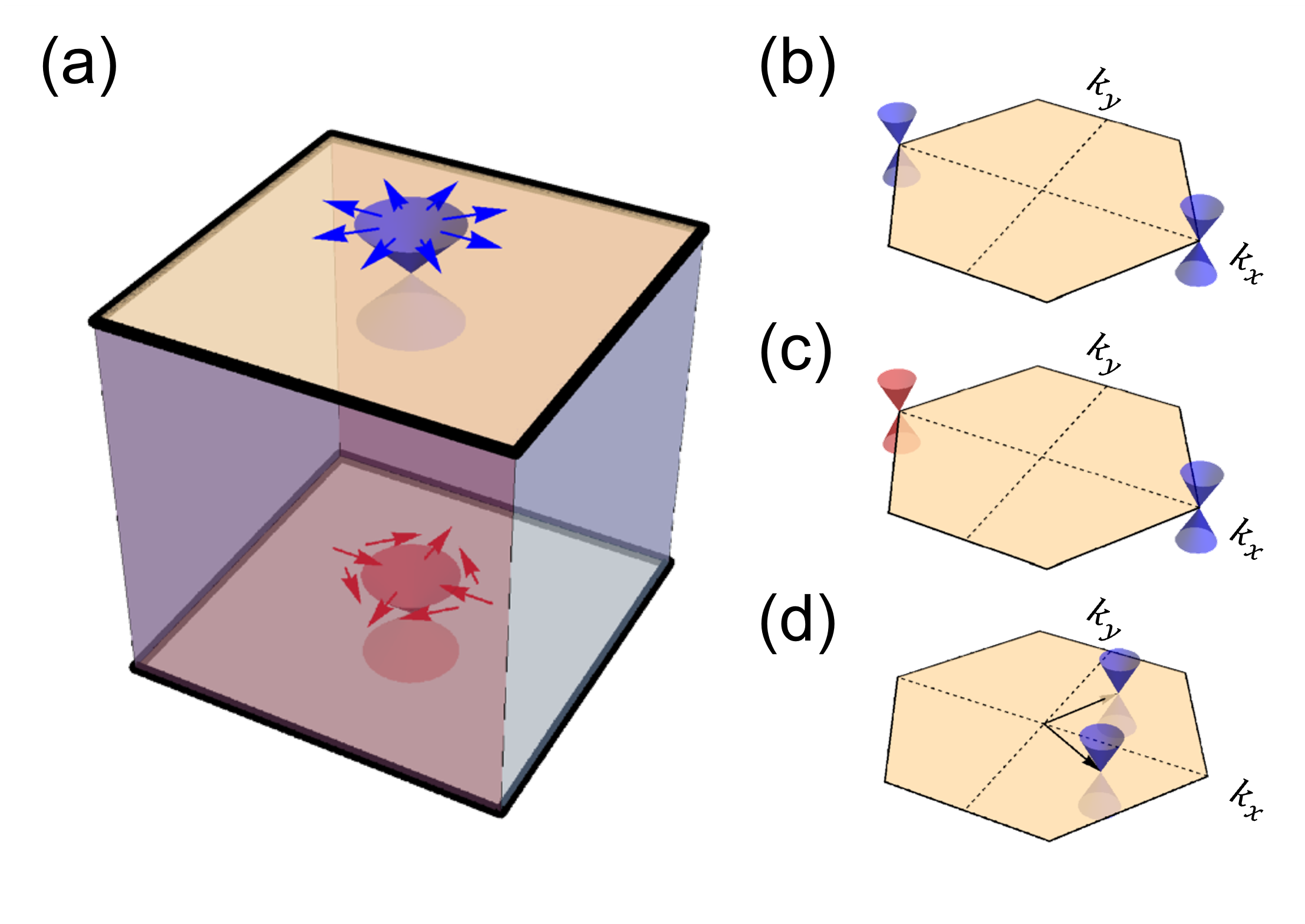}
\caption{{\bf Nodes and vorticity of 2D Dirac fermions in solids.}
(a) Dirac fermions on the surfaces of TRS-protected 3D TI. The red (blue) color represents the $+1$ ($-1$) vorticity. (b) Dirac fermions between two 2D bands with nonzero Euler number.  (c) Dirac fermions between two 2D bands with zero Euler number (e.g. graphene). When the \textit{physical} (\textit{effective}) TRS exists, the relative positions of the nodes are fixed and connected by TRS as in (b) and (c). (d) When joint symmetry $PT$ or $C_{2z}T$ that is local in $k$-space exists whereas $T$ does not. The positions of the nodes are not restricted by symmetry.
}
\label{fig_anomaly} 
\end{figure}

\textit{Introduction}---Symmetry protected topological phases, such as topological insulators (TIs) and topological superconductors (TSCs) are characterized by robust gapless boundary states protected by stable bulk topology~\cite{bernevig2006quantum,moore2010birth,ludwig2015topological}. 
These delocalized boundary modes can remain conducting even in the presence of disorder, evading the Anderson localization which suppresses charge or heat transport~\cite{ryu2010topological,ludwig2015topological,anderson1958absence,evers2008anderson}. An important feature of such topological boundary states is the violation of the fermion doubling theorem that predicts the zero total vorticity (or chirality) of Dirac fermions in the bulk Brillouin zone (BZ) of a periodic lattice~\cite{nielsen1981no}. At a boundary, where periodicity is lost, this constraint can be circumvented~\cite{fu2007topological,moore2007topological,roy2009z,teo2010topological}. For instance, as described in Figure~\ref{fig_anomaly}(a), a single Dirac fermion can appear on the surface of a three-dimensional (3D) time-reversal symmetry (TRS)-protected TI, with the anomaly resolved by the opposite surface carrying the opposite vorticity. Because a Dirac fermion can be gapped only by pair annihilation with another Dirac fermion of opposite vorticity~\cite{prodan2011disordered}, the TI surface remains gapless under disorder as long as TRS is preserved and the two surfaces are well-separated.

Beyond conventional TIs with stable topology, fragile and delicate topological phases have been identified, in which the topology of a group of bands depends on additional bands~\cite{po2018fragile,song2020twisted,hwang2019fragile,nelson2021multicellularity,lim2023real}. 
A representative fragile topological band structure is provided by two-dimensional (2D) Euler bands characterized by a nonzero Euler invariant $e_2$~\cite{ahn2019failure,bouhon2020non,bouhon2020geometric,wu2019non,jiang2021experimental}. Two bands with nonzero $e_2$, called the Euler bands, necessarily contain band crossings whose total vorticity equals $2e_2$, implying a violation of the fermion-doubling constraint with the bulk band structure~\cite{ahn2019failure} as illustrated in Figure~\ref{fig_anomaly}(b). 
This makes the Euler bands an intriguing platform to study disorder effects on the bands with nonzero total vorticity [Figure~\ref{fig_anomaly}(b)], distinct from the conventional 2D Dirac semimetals such as graphene with zero total vorticity [Figure~\ref{fig_anomaly}(c)]. While disorder in Euler models has been studied mainly focusing on disorder-driven phase transitions under disorder that preserves the relevant symmetries on average~\cite{jankowski2023disorder,wang2024anderson,unal2020topological}, their weak-localization (WL)/weak-antilocalization (WAL) behavior has received much less attention.

In 2D crystalline systems where the Bloch Hamiltonian can be made globally real due to an antiunitary space-time inversion symmetry $I_{ST}$ that is local in momentum space and satisfies $I_{ST}^2=+1$, one can define an integer-valued Euler number $e_2$ for a pair of isolated real Bloch bands~\cite{ahn2019failure,jiang2021experimental}. 
$I_{ST}$ can appear in the form of $PT$ in spinless systems or $C_{2z}T$ in both spinful ($T^2=-1$) and spinless ($T^2=+1$) systems, where $C_{2z}$, $P$, and $T$ indicate two-fold rotation about the $z$-axis, spatial inversion, and TRS, respectively.

One representative spinful electronic system hosting Euler bands is a magnetic Euler insulator, which can be obtained by introducing in-plane magnetism to a $C_{2z}$ symmetric quantum spin Hall insulator~\cite{lee2025euler}. 
In magnetic Euler bands, where in-plane magnetization and spin-orbit coupling exist simultaneously~\cite{lee2025euler}, \textit{physical} TRS is explicilty broken. Thus, it is expected that the system should belong to the unitary class (class A) without logarithmic WL corrections. 
Interestingly, however, we show that this naive expectation fails when $P$ symmetry is additionally present. In the magnetic Euler bands with extra $P$ symmetry, $P$ combined with $C_{2z}T$ symmetry generates an \textit{effective} TRS
\begin{align}
T^*=P(C_{2z}T)=M_zT~,
\end{align}
in which $(T^*)^2=+1$ and $M_z$ indicates the mirror symmetry with respect to the 2D material plane. 

Using a magnetic Euler model on the kagome lattice, we identify the symmetry-allowed disorder channels and evaluate the WL/WAL corrections diagrammatically, considering the Cooperon contributions~\cite{hikami1980spin,mccann2006weak}. We further show that, despite the distinct nodal-vorticity patterns between graphene and the magnetic Euler bands, the Cooperon correction is equivalent in these two systems under a suitable basis change~\cite{mccann2006weak}. Magnetic Euler bands therefore provide a concrete setting in which crystalline symmetry reshapes the universality class under disorder: within the weak-disorder regime considered here, the decisive ingredient is the relevant \textit{physical} or \textit{effective} TRS, not the relative vorticity between the Dirac nodes.

\begin{figure}[t]
\centering
\includegraphics[width=0.5\textwidth]{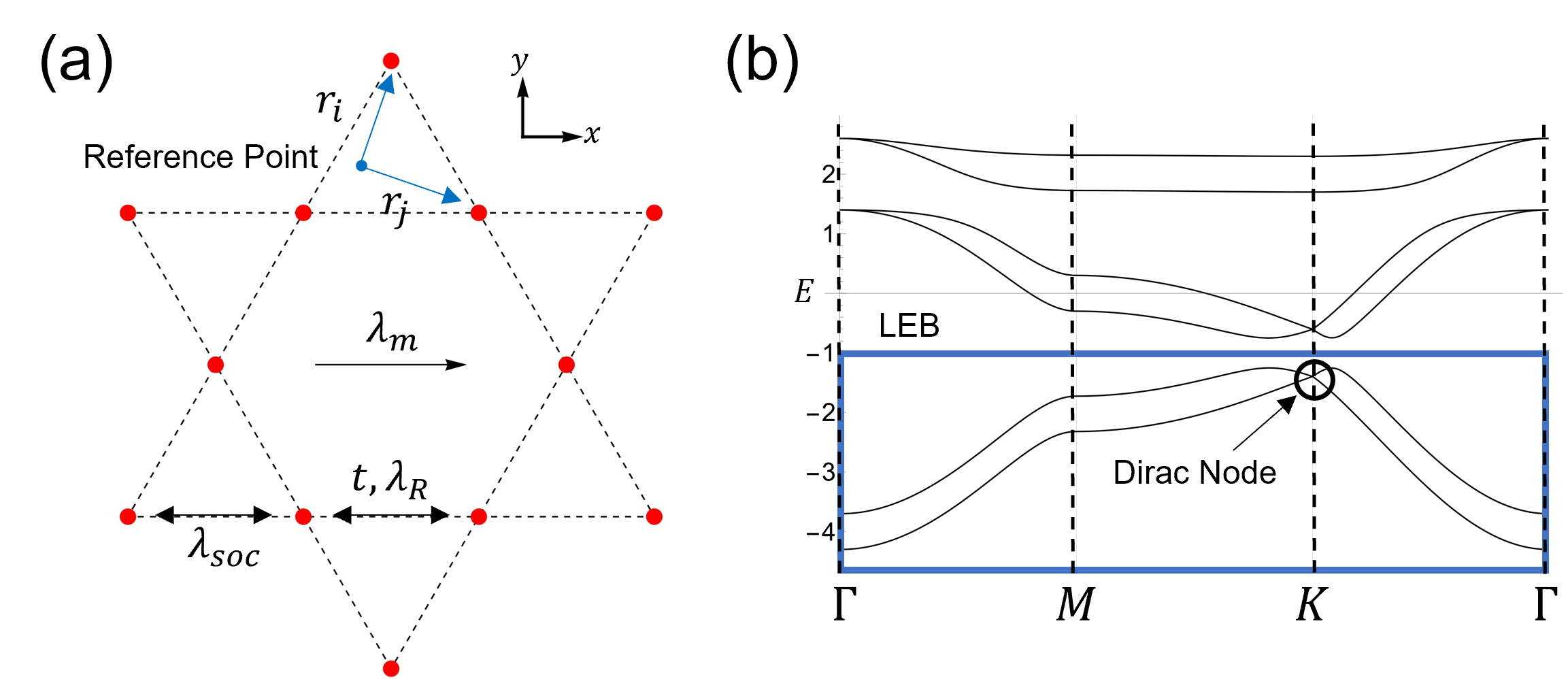}
\caption{{\bf Kagome lattice model with spin-orbit coupling and in-plane magnetism.} (a) Kagome lattice used to build the model Hamiltonian. 
Here, $I_{ST}=C_{2z}T$ is not broken by the addition of Rashba effect ($\lambda_R$) and in-plane magnetization ($\lambda_m$). The blue dot denotes the reference point in unit cell. For hopping from $j$ to $i$, $\nu_{ij}=+1 (-1)$ when $z$ component of $r_i\times r_j$ is positive (negative). (b) The band structure for $t=1,~\lambda_{soc}=0.1,~\lambda_R=0,~\textrm{and}~\lambda_m=0.3$ in arbitrary energy unit. LEB (blue box) possesses two Dirac nodes at the $K$ and $K'$ point with the same vorticity +1, respectively.}
\label{fig_spinfulKagome}
\end{figure}

\textit{Model construction}---To demonstrate our theory, we construct a magnetic Euler band model, starting with a spin-orbit coupled kagome lattice~\cite{lee2025euler}. In addition to intrinsic spin-orbit interactions, we consider Rashba-type spin-orbit coupling and Zeeman coupling to an in-plane magnetization,

\begin{widetext}
\begin{equation}
H=-\sum_{\langle ij\rangle}t c_i^{\dagger}c_j-\sum_{\langle ij\rangle}i\lambda_{soc}\nu_{ij}c_i^{\dagger}\sigma_zc_j+\sum_{\langle ij\rangle}i\lambda_{R}c_i^{\dagger}(\bm{\sigma}\times\hat{\bm{d}}_{ij})_zc_j+\sum_i\lambda_mc_i^{\dagger}\sigma_xc_i~,
\label{eq_fullHam}
\end{equation}
\end{widetext}
where $c_i = (c_{i\uparrow}, c_{i\downarrow})^T$ is a spinor of electron annihilation operator, $t$ represents the hopping strength between nearest-neighbor sites $(i,~j)$ and $\lambda_{soc}$ is the spin-orbit coupling strength which originates from the presence of the third neighboring sublattice other than $i$ and $j$. To fix $\nu_{ij}=\pm1$, consider an arbitrary reference point in the unit cell (the blue point in Figure~\ref{fig_spinfulKagome}(a)). Then, let $r_i$ and $r_j$ be the positions of lattice site $i$ and $j$ with respect to the reference point. If $(r_i\times r_j)_z>0$, we set $\nu_{ij}=+1$ and vice versa. $\lambda_R$ is the Rashba spin-orbit coupling strength, $\bm{\sigma}=(\sigma_x,\sigma_y,~\sigma_z)$ are Pauli matrices representing physical spin, and $\hat{\bm{d}}_{ij}$ is a unit vector along the bond where an electron hops from site $j$ to $i$. $\lambda_m\propto B$ is the Zeeman energy with $\bm{B}=B\hat{x}$ arising from the exchange coupling to the in-plane magnetization as shown in Figure~\ref{fig_spinfulKagome}(a). 
A notable property of this model is that in-plane magnetism represented by nonzero $\lambda_m$ breaks $C_{2z}$ and $T$ symmetries separately, but preserves the combined space-time inversion symmetry $I_{ST}=C_{2z}T$, so that Euler bands remain well-defined. Additionally, Rashba spin-orbit coupling breaks the inversion symmetry, thus setting $\lambda_R=0$ restores $P$ symmetry. This determines whether the \textit{effective} TRS $T^*\equiv P(C_{2z}T)$ exists, and therefore whether WL can appear.

A typical band structure of the magnetic Euler band model with $\lambda_R=0$ and $\lambda_m\neq0$ is composed of three groups of isolated bands as in Figure~\ref{fig_spinfulKagome}(b). The uppermost and lowermost bands exhibit nontrivial Euler numbers, while the middle bands have $e_2=0$. In particular, the lowermost Euler bands (LEB)---the blue box in Figure~\ref{fig_spinfulKagome}(b)--- with $e_2=1$ possess two Dirac nodes with the same vorticity $+1$ at the $K$ and $K'$ points, related by \textit{effective} TRS $T^*$, as in Figure~\ref{fig_anomaly}(b). On the other hand, when $\lambda_R\neq0$, the two nodes are not confined at $K$ and $K'$ [Figure~\ref{fig_anomaly}(d)], since neither \textit{physical} nor \textit{effective} TRS exists.

For the system that exhibits the \textit{effective} TRS $T^*$ ($\lambda_R=0$), we project the total $6\times6$ Hamiltonian onto the two nodes of the LEB and obtain the $4\times4$ effective low-energy Hamiltonian given by
\begin{align}
H_{\textrm{LEB}}=v\Pi_z\otimes\bm{k}\cdot\bm{\rho}~,
\label{eq_case3_LEB}
\end{align}
where $v,~\bm{k}=(k_x,k_y)$ represent the Fermi velocity and relative momenta with respect to the nodes, respectively. $\bm{\Pi}=(\Pi_x,\Pi_y,\Pi_z)$ and $\bm{\rho}=(\rho_x,\rho_y,\rho_z)$ are Pauli matrices acting in the valley space ($K,~K'$) and in the pseudospin space describing the two bands crossing at each valley, respectively. The basis choice and detailed derivation of Eq.~(\ref{eq_case3_LEB}) are provided in the Supplemental Material (SM)~\cite{supplement}. We note that $H_{\textrm{LEB}}$ must be distinguished from the effective Hamiltonian of graphene, which exhibits two valleys with opposite vorticities. One might argue that the sign of vorticities can be altered through basis transformations (e.g., by permuting the basis order), rendering it meaningless. However, to properly account for the Euler band topology, the orientation of eigenstates must be kept consistent across the entire BZ~\cite{ahn2019failure}. To ensure this condition, we fix the orientation of the Hamiltonian.

\textit{Symmetry classification of spinful Euler bands}---The symmetry of Eq.~(\ref{eq_fullHam}) is controlled by the Rashba coupling $\lambda_R$ and the in-plane Zeeman field $\lambda_m$, which gives three possible cases as follows. 
For $\lambda_R\neq0$ and $\lambda_m=0$, $P$ symmetry is broken, while $C_{2z}$ symmetry and the \textit{physical} TRS $T$ are preserved. For $\lambda_R\neq0$ and $\lambda\neq0$, both $P$ and $T$ symmetries are broken, while the Euler bands are protected by the joint symmetry $C_{2z}T$. For $\lambda_R=0$ and $\lambda_m\neq0$, the in-plane field breaks $T$ symmetry. However, an \textit{effective} TRS $T^*=M_zT$ emerges due to the restored $P$ symmetry. 
The last case corresponds to spinful magnetic Euler bands with broken \textit{physical} TRS but preserved \textit{effective} TRS. 
In the following, we focus on the last case because the WL/WAL behaivors of other two cases can be understood straightforwardly based on the \textit{physical} symmetry of the system [see the SM].

\textit{Symmetry analysis}---
Let us examine the disorder effect on the LEB described by Eq.~(\ref{eq_case3_LEB}). The two Dirac nodes in the LEB act as sources of a $\pi$-Berry phase. When each valley of the LEB is considered independently, the $\pi$-Berry phase suppresses electron backscattering, thereby contributing to WAL corrections to the conductivity~\cite{ando1998berry}, rather than conventional WL or the lack of quantum corrections. To investigate this WAL behavior under symmetry constraints, we adopt a basis convention that aligns our valley-pseudospin labeling with Ref.~\cite{mccann2006weak}. As the orientation of the basis is globally fixed in Euler bands, the relative vorticities are different from graphene case. To highlight this vorticity difference, we reorder the basis in the second valley of the LEB so that the basis order changes from $\ket{u_K}_1,\ket{u_K}_2,\ket{u_{K'}}_1,\ket{u_{K'}}_2$ to $\ket{u_K}_1,\ket{u_K}_2,\ket{u_{K'}}_2,\ket{u_{K'}}_1$. This basis exchange does not alter physical observables or disorder operators, while it enables a direct comparison with the calculation of WL/WAL corrections in graphene reported in Ref.~\cite{mccann2006weak}. In this convention, the Hamiltonians for the two systems are written as
\begin{align}
&H_{\textrm{graphene}}=v
\Pi_z\otimes\bm{k}\cdot\bm{\rho}~,
\label{eq_lowEHam_graphene}
\\
&\widetilde{H}_{\textrm{LEB}}=v\left[\frac{1}{2}(\mathbbm{1}_2+\Pi_z)\otimes\bm{k}\cdot\bm{\rho}-\frac{1}{2}(\mathbbm{1}_2-\Pi_z)\otimes(\bm{k}\cdot\bm{\rho})^\textrm{T}\right]~.
\label{eq_lowEHam_LEB}
\end{align}

Let us introduce sets of matrices for $\widetilde{H}_{\textrm{LEB}}$ that are convenient for dealing with disorder potentials~\cite{mccann2006weak}:
\begin{align}
\label{eq_dispot}
 \bm{\Sigma}&=(\Sigma_x,\Sigma_y,\Sigma_z)\equiv(\Pi_z\otimes\rho_x, \Pi_z\otimes\rho_y, \Pi_0\otimes\rho_z)~,
 \\
\bm{\Lambda}&=(\Lambda_x,\Lambda_y,\Lambda_z)\equiv(\Pi_x\otimes\rho_z, \Pi_y\otimes\rho_z, \Pi_z\otimes\rho_0)~,
\end{align}
and $\Sigma_0=\Lambda_0\equiv\mathbb{1}_4$ the $4\times4$ identity matrix.
The $\Sigma$ and $\Lambda$ matrices form two independent SU(2) subgroups of U(4) and provide a convenient basis for describing disorder potentials acting on the LEB. Under the \textrm{effective} TRS $T^*_{\textrm{LEB}}$, which is the projection of $T^*$ symmetry onto the LEB, these matrices transform as
\begin{align}
\label{eq_distrf}
&T_{\textrm{LEB}}^*\Sigma_i(T_{\textrm{LEB}}^*)^{-1}=(-1)^{\delta_{i,x}}\Sigma_i~,
\\
&T_{\textrm{LEB}}^*\Lambda_i(T_{\textrm{LEB}}^*)^{-1}=(-1)^{\delta_{i,z}}\Lambda_i~,
\end{align}
indicating that only $\Sigma_x$ and $\Lambda_z$ change sign, while the other components remain invariant. Here, $\delta_{i,j}$ is a Kronecker delta. Accordingly, a disorder potential $\Sigma_i\Lambda_j$ is $T^*_{\textrm{LEB}}$-invariant when the total number of sign-changing matrices is even, namely when both or neither of $\Sigma_x$ and $\Lambda_z$ appear in $\Sigma_i\Lambda_j$.

Using $\Sigma_i\Lambda_j$ as a matrix basis, all 16 disorder potentials can be categorized into three types: scalar disorder potential $V_{\textrm{sc}}\propto\Sigma_0\Lambda_0$ leaving valley and pseudospin degrees of freedom unchanged, intravalley disorder potentials $V_{\textrm{intra}}\propto(\Sigma_x\Lambda_z,\Sigma_x\Lambda_0,\Sigma_y\Lambda_z,\Sigma_y\Lambda_0,\Sigma_z\Lambda_z,\Sigma_z\Lambda_0,\Sigma_0\Lambda_z)$ that induce scattering between different pseudospin states within a single valley, and intervalley disorder potentials $V_{\textrm{inter}}\propto(\Sigma_x\Lambda_x,\Sigma_x\Lambda_y,\Sigma_y\Lambda_x,\Sigma_y\Lambda_y,\Sigma_z\Lambda_x,\Sigma_z\Lambda_y,\Sigma_0\Lambda_x,\Sigma_0\Lambda_y)$ that scatter electrons from one valley to the other. For magnetic Euler bands with $\lambda_R=0$, we restrict our discussion to the disorder channels that are even under $T_{\textrm{LEB}}^*$, $P_{\textrm{LEB}}$ and $(C_{2z}T)_{\textrm{LEB}}$, where $P_{\textrm{LEB}}$ and $(C_{2z}T)_{\textrm{LEB}}$ are the projection of $P$ and $C_{2z}T$ symmetries onto the LEB. (We require $C_{2z}T$ symmetry to be conserved because it is essential for the system to have the stable Euler bands. Furthermore, $P$ symmetry is kept for the construction of the effective TRS $T_{\textrm{LEB}}^*$. The matrix representations of $T_{\textrm{LEB}}^*$, $P_{\textrm{LEB}}$ and $\left(C_{2z}T\right)_{\textrm{LEB}}$ are explicitly shown in SM.) The symmetry invariant subset of disorder potentials includes $\Sigma_x\Lambda_z$ and $\Sigma_y\Lambda_0$ for intravalley scattering and $\Sigma_y\Lambda_y$ $\Sigma_z\Lambda_x$ and $\Sigma_0\Lambda_y$ for intervalley scattering. Then the disordered LEB system is described by~\cite{mccann2006weak}
\begin{align}
\label{eq_htot}
H_{\textrm{tot}}=\widetilde{H}_{\textrm{LEB}}+\sum_{i,j}u_{ij}\Sigma_i\Lambda_j~,
\end{align}
where $u_{ij}$ is the disorder potential strength that induces the scattering rate $\tau_{ij}^{-1}\equiv\pi N_0u_{ij}^2$ with the density of states at the Fermi energy $N_0=E_F/2\pi v^2$. Note that we keep only $\Sigma_i\Lambda_j$s that are invariant under $T_{\textrm{LEB}}^*$, $P_{\textrm{LEB}}$, $\left(C_{2z}T\right)_{\textrm{LEB}}$. For simplicity, we consider the case where different disorder potentials are treated independently and the scalar disorder dominates the elastic scattering, i.e., $|u_{00}|\gg|u_{ij}|$ for all other channels.

Let us first consider the case with scalar disorder only. Then the scattering is confined to a single valley, allowing each valley to be treated independently. A single valley Hamiltonian takes the form $\pm(k_x\rho_x+k_y\rho_y)+u_{00}\rho_0$, which admits a local TRS acting within each valley, $T_{\textrm{SV}}=i\mathbb{1}\otimes\rho_yK$, where $K$ denotes complex conjugation. We note that $T_{\textrm{SV}}$ is a valley-local antiunitary symmetry of the single valley Dirac Hamiltonian, distinct from the \textit{physical} TRS. Since $T_{\textrm{SV}}^2=-1$, the quantum correction to the conductivity is expected to be positive, indicating WAL. When we introduce general intravalley disorder potentials, $T_{\textrm{SV}}$ is broken ($T_{\textrm{SV}}V_{\textrm{intra}}T_{\textrm{SV}}^{-1}\neq V_{\textrm{intra}}$). Then the system belongs to the unitary class (class A)~\cite{ryu2010topological}, in which the logarithmic quantum correction to the conductivity should vanish. Finally, when intervalley disorder potentials are turned on, scattering between the two valleys becomes possible, and the relevant antiunitary symmetry is no longer represented by $T_{\textrm{SV}}$. Instead, the \textrm{effective} TRS of the LEB, $T_{\textrm{LEB}}^*$, becomes the relevant symmetry constraint for quantum interference. Because $(T_{\textrm{LEB}}^*)^2=+1$, it leads to WL corrections to the conductivity. We therefore conclude that the crossover from the symplectic class (class A\rom{2}) to the orthogonal class (class A\rom{1}) and the unitary class (class A) occurs when intervalley and intravalley disorder potentials are introduced, respectively.

We note that these results are the same as the localization properties of graphene~\cite{mccann2006weak,tikhonenko2009transition}, with an important difference that the \textit{physical} TRS triggers the crossover in graphene instead of the \textit{effective} TRS. Although two systems show distinct vorticity configurations, we expect that the presence of the \textit{physical} or \textit{effective} TRS should govern the quantum correction to the conductivity due to disorder, not the relative vorticities between Dirac nodes. Below we confirm this using diagrammatic calculations.

\begin{figure}[t]
\centering
\includegraphics[width=\linewidth]{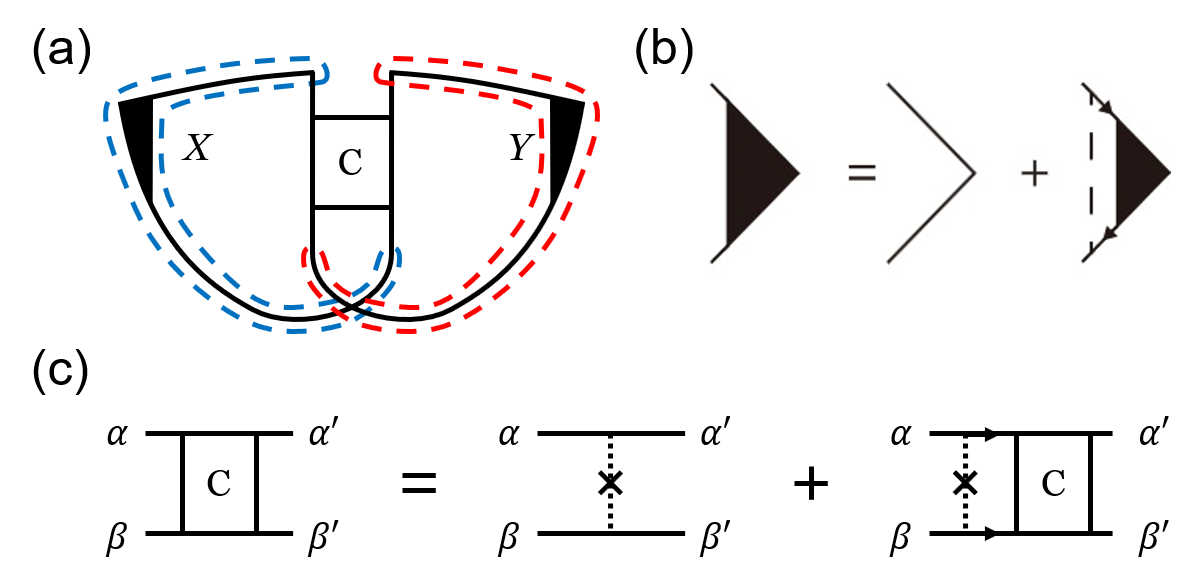}
\caption{{\bf Feynman diagrams for the quantum correction to the conductivity.}
(a) Diagrammatic representation of the current-current correlation function that gives the conductivity correction by disorder $\delta g$. $X$ (blue) and $Y$ (red) denote left and right legs connected to the Cooperon, respectively. (b) Vertex correction due to low-angle backscattering. (c) Bethe-Salpeter diagram for the Cooperon.}
\label{fig_manybody} 
\end{figure}

\textit{Diagrammatic calculations}---We examine the influence of symmetry on the WL/WAL behavior by evaluating the disorder-induced quantum corrections $\delta g$ for the LEB using the standard retarded-advanced Green's function formulation for the DC conductivity in the weak-disorder regime $(E_F\tau_{00}\gg1)$, including the ladder vertex correction and the Cooperon contribution shown in Figure~\ref{fig_manybody}(b), (c)~\cite{hikami1980spin} in the weak-disorder and DC limit.

For scalar disorder potential $V_{\textrm{sc}}$, the valleys are effectively decoupled and we recover a positive logarithmic correction characteristic of WAL. Addition of intravalley disorder potentials $V_{\textrm{intra}}$ leads to the saturation of $\delta g$ without the diverging log correction. On the other hand, when we include intervalley disorder potentials, two valleys become connected by $V_{\textrm{inter}}$,
\begin{align}
\Sigma_y\Lambda_y,~\Sigma_z\Lambda_x,~\Lambda_y
\label{eq_disLEB}
\end{align}
satisfying $SV_{\textrm{inter}}S^{-1}=V_{\textrm{inter}}$, where $S=T_{\textrm{LEB}}^*$, $P_{\textrm{LEB}}$, $(C_{2z}T)_{\textrm{LEB}}$. The direct calculation of Figure~\ref{fig_manybody}(c) using these disorder potentials produces the WL correction even in the presence of in-plane magnetization, which is a consequence of the combination of $P$ and $C_{2z}T$ symmetries. The final results are summarized in Table~\ref{tb_1}. See the SM for detailed calculations~\cite{supplement}.

We further compare the LEB and graphene to see whether the relative vorticities between Dirac nodes affect the quantum correction to the conductivity due to disorder. Namely, in graphene, valleys at $K$ and $K'$ have opposite vorticities (Figure~\ref{fig_anomaly}(c)), whereas in the Euler band model they possess the same vorticity (Figure~\ref{fig_anomaly}(b)). It is well known that in graphene, intervalley disorder potentials give WL correction to the conductivity due to \textit{physical} TRS~\cite{mccann2006weak,khveshchenko2006electron,bardarson2007one,suzuura2002crossover,ostrovsky2006electron}. On the other hand, in the LEB without \textit{physical} TRS, intervalley disorder potentials give WL correction to the conductivity due to the \textit{effective} TRS. Despite the difference in the relative vorticity configuration, both systems yield the same type of quantum correction, WL. This indicates that the localization behavior is governed by the presence or absence of \textit{physical} or \textit{effective} TRS, rather than the relative vorticity between the Dirac nodes.

\begin{table}[t]
\caption{{\bf Quantum corrections $\delta g$ to the conductivity due to disorder in LEB.} We calculated $\delta g$ from the Cooperon diagram of Fig.~\ref{fig_manybody}(a) for each disorder potential. The results from symmetry analysis are also shown. Here, $\tau_{\phi}^{-1}$ represents the inelastic scattering rate; $\tau_{00}^{-1}$ is the scattering rate due to scalar disorder potentials;  $\tau_{\textrm{intra}}^{-1}$ and $\tau_{\textrm{inter}}^{-1}$ are the sum of the scattering rates due to intravalley and intervalley disorder potentials, respectively. The total scattering rate $\tau^{-1}$ is defined by $\tau^{-1}\equiv\tau_{00}^{-1}+\tau_{\textrm{intra}}^{-1}$ for the $V_{\textrm{sc}}+V_{\textrm{intra}}$ case and by $\tau^{-1}\equiv\tau_{00}^{-1}+\tau_{\textrm{inter}}^{-1}$ for the $V_{\textrm{sc}}+V_{\textrm{inter}}$ case. At low temperature, $\tau_{\phi}\rightarrow\infty$, making the logarithmic term $\ln(\tau_{\phi}/\tau_{00})$ and $\ln(\tau_{\phi}/\tau)$ divergent. However, $\delta g$ for the scalar+intravalley disorder potential case saturate to finite constant as $T\rightarrow0$. We note that including all three types of disorder potentials ($V_{\textrm{sc}}$, $V_{\textrm{intra}}$, and $V_{\textrm{inter}}$) also leads to the orthogonal class as in the $V_{\textrm{sc}}+V_{\textrm{inter}}$ case.}
\begin{tabular}{c|c|c|c|c}
\hline\hline
\multicolumn{2}{c|}{System}&$\delta g /\left(\frac{e^2}{\pi^2}\right)$&TRS&Symmetry class
\\
\hline
\multirow{3}{*}{\shortstack{Disorder\\potential}}&
$V_{\textrm{sc}}$&$2\ln\left(\tau_{\phi}/\tau_{00}\right)$&$T_{\textrm{sv}}$&\textrm{Symplectic}
\\
&$V_{\textrm{sc}}+V_{\textrm{intra}}$&$2\ln\left(\frac{2\tau/\tau_{\textrm{intra}}+1}{2\tau/\tau_{\textrm{intra}}+\tau/\tau_{\phi}}\right)$&$-$&\textrm{Unitary}
\\
&$V_{\textrm{sc}}+V_{\textrm{inter}}$&$-\ln\left(\tau\tau_{\phi}/\tau_{\textrm{inter}}^2\right)$&$T_{\textrm{LEB}}^*$&\textrm{Orthogonal}
\\
\hline\hline
\end{tabular}
\label{tb_1}
\end{table}

To understand this result, let us describe some details of diagrammatic calculations starting from the effective Hamiltonians near the Dirac nodes. The key distinction between graphene and the Euler band model originates from their Hamiltonians in Eq.~(\ref{eq_lowEHam_graphene}) and (\ref{eq_lowEHam_LEB}), respectively, which encode opposite and identical relative vorticities, respectively, while the disorder potentials are taken identical, $V_{ij}=u_{ij}\Sigma_i\Lambda_j$ as in Ref.~\cite{mccann2006weak}. Since two Hamiltonians are connected by a unitary transformation $U_{\textrm{exch}}\equiv\mathbbm{1}_2\oplus\rho_x$ exchanging the pseudospin direction of the second valley, the corresponding Green's function and velocity operator are also transformed by $U_{\textrm{exch}}$. This implies that the legs $X$ and $Y$ appearing in Figure~\ref{fig_manybody}(a) of graphene and the LEB are connected by $U_{\textrm{exch}}$,
\begin{align}
\label{eq_Utrf}
U_{\textrm{exch}}X^{\textrm{graphene}}U_{\textrm{exch}}^{\dag}=X^{\textrm{LEB}}~,~U_{\textrm{exch}}Y^{\textrm{graphene}}U_{\textrm{exch}}^{\dag}=Y^{\textrm{LEB}}~,
\end{align}
since $X$ and $Y$ are a combination of the retarded and advanced Green's function and the velocity operator,
\begin{align}
&X_j(\bm{k})\equiv G_R(\bm{k})\tilde{v}_jG_A(\bm{k})~,\nonumber
\\
&Y_j(\bm{k})\equiv G_A(-\bm{k})\tilde{v}_jG_R(-\bm{k})~,    
\end{align}
where $\tilde{v}_{j}$ is the vertex-corrected velocity operator and $j=x, y$.

Then we consider disorder potentials for the Cooperon part, $C_{\alpha'\beta',\alpha\beta}$. Intervalley disorder potentials that satisfy the TRS $T_{\textrm{graphene}}=\Sigma_y\Lambda_yK$ in graphene are $\Sigma_x\Lambda_x,\Sigma_x\Lambda_y,\Sigma_y\Lambda_x,\Sigma_y\Lambda_y,\Sigma_z\Lambda_x,\Sigma_z\Lambda_y$. [Among these intervalley disorder potentials, only $V_{\textrm{graphene,inter}}=\Sigma_x\Lambda_y$, $\Sigma_y\Lambda_y$, $\Sigma_z\Lambda_x$ preserve $P_{\textrm{graphene}}$ and $(C_{2z}T)_{\textrm{graphene}}$ symmetries of graphene, analogous to the LEB.] When we perform a unitary transformation with $U_{\textrm{exch}}$, we find that these graphene disorder potentials become equivalent to $T^*_{\textrm{LEB}}$-invariant disorder potentials in Eq.~(\ref{eq_disLEB}) of the LEB up to sign, $U_{\textrm{exch}}V_{\textrm{graph,inter}}U_{\textrm{exch}}^{\dagger}=\pm V_{\textrm{LEB,inter}}.$ However, these signs are not important when calculating the quantum correction to the conductivity due to disorder in the Born approximation, since disorder potentials sharing the same sign always appear an even number of times in the Cooperon. Thus, we infer that Cooperons have the same form in graphene and the LEB under the exchange of the pseudospin indices of the second valley. Finally, in the diagrammatic expression of $\delta g$, all components of pseudospin degrees of freedom equally contribute. This leads to the conclusion that Cooperon correction diagrams are the same for graphene and the Euler insulator. See the SM for more details~\cite{supplement}.
    
\textit{Discussion}---We have studied the disorder-induced correction to the conductivity in Euler band models by classifying the symmetry classes of 2D electronic Euler bands and predicting their behavior based on symmetry considerations. We confirmed our symmetry-based analyses through diagrammatic calculations by using a magnetic Euler insulator model introduced recently~\cite{lee2025euler}, and incorporating Rashba-type spin-orbit coupling to control inversion symmetry breaking~\cite{supplement}.

Our results highlight a striking interplay between crystalline symmetry and disorder classification. Typically, when spinful fermions preserve a TRS with $T^2=-1$, they exhibit WAL belonging to the symplectic class~\cite{ryu2010topological,evers2008anderson}. In our magnetic Euler bands, however, the in-plane magnetism explicitly breaks the \textit{physical} TRS, which would conventionally drive the system into the unitary class, whereas $I_{\textrm{ST}}=C_{2z}T$ is intact. Remarkably, the presence of inversion symmetry prevents this by facilitating the emergence of the \textit{effective} TRS, $T^*=M_zT$. Unlike the \textit{physical} TRS, this \textit{effective} TRS squares to +1 even in the presence of spin-orbit coupling. This leads to the counter-intuitive result that a spinful, magnetic system exhibits WL characteristic of the orthogonal class. This demonstrates that \textit{effective} crystalline symmetries can supersede the conventional localization behavior naively predicted based on the \textit{physical} symmetry only.

We conclude that the quantum correction to the conductivity due to disorder is determined by either \textit{physical} TRS or \textit{effective} TRS of the system, irrespective of the relative vorticities between Dirac points. 
However, we note that this conclusion is based on the diagrammatic calculations with the Born approximation. Examining the influence of the higher order scattering processes is beyond the scope of the current paper, which we leave for future study. 

\begin{acknowledgments}
D.Y.K., S.L., B.-J.Y. were supported by Samsung
Science and Technology Foundation under Projects No.~SSTF-BA2601-02, National Research Foundation of Korea
(NRF) funded by the Korean government (MSIT),
Grants No.~RS-2021-NR060087 and No.~RS-2025-00562579,
Global Research Development-Center (GRDC) Cooperative
Hub Program through the NRF funded by the MSIT,
Grant No.~RS-2023-00258359, and Global-LAMP program of
the NRF funded by the Ministry of Education, Grant No.~RS-2023-00301976. A.F. was supported by JSPS KAKENHI, Grant No.~JP19K03680, and JST CREST, Grant No.~JPMJCR19T2.
\end{acknowledgments}

\bibliographystyle{apsrev4-2}
\bibliography{WL}

\clearpage
\onecolumngrid
\begin{center}
\textbf{\large Supplemental Material for ``Weak localization in magnetic Euler bands"}
\end{center}
\begin{center}
\text{Doh-Young Kim,$^{1,~2,~3,~*}$ Seung Hun Lee,$^{1,~2,~3,~*}$ Akira Furusaki,$^{4,~\dag}$ and Bohm-Jung Yang$^{1,~2,~3,~\ddag}$} \\
$^{1}$ \textit{Department of Physics and Astronomy, Seoul National University, Seoul 08826, Korea} \\
$^{2}$ \textit{Center for Theoretical Physics (CTP), Seoul National University, Seoul 08826, Korea} \\
$^{3}$ \textit{Institute of Applied Physics, Seoul National University, Seoul 08826, Korea} \\
$^{4}$ \textit{RIKEN Center for Emergent Matter Science, Wako, Saitama, 351-0198, Japan}
\end{center}

\setcounter{section}{0}
\setcounter{figure}{0}
\setcounter{equation}{0}
\renewcommand{\thefigure}{S\arabic{figure}}
\renewcommand{\theequation}{S\arabic{equation}}
\renewcommand{\thesection}{S\arabic{section}}

\section{Orientation Matching of the Euler Bands}
\label{supsec_OrMatch}
In this section, we discuss the orientation matching of the eigenstates, which is necessary for the construction of the Euler bands~\cite{ahn2019failure}. For the Hamiltonian describing the Euler model,
\begin{align}
\label{eq_supHamil}
H=-\sum_{\langle ij\rangle}t c_i^{\dagger}c_j-\sum_{\langle ij\rangle}i\lambda_{soc}\nu_{ij}c_i^{\dagger}\sigma_zc_j+\sum_{\langle ij\rangle}i\lambda_{R}c_i^{\dagger}(\bm{\sigma}\times\hat{\bm{d}}_{ij})_zc_j+\sum_i\lambda_mc_i^{\dagger}\sigma_xc_i~,\quad c_i\equiv
\begin{pmatrix}
c_{i\uparrow}\\
c_{i\downarrow}
\end{pmatrix}~,
\end{align}
the symmetries are controlled by the parameters $\lambda_R$ describing the Rashba-type coupling and $\lambda_m$ describing the effective Zeeman coupling induced by an in-plane magnetization. Here, $c_i = (c_{i\uparrow}, c_{i\downarrow})^T$ is a spinor of electron annihilation operator. Because the Hamiltonian Eq.~(\ref{eq_supHamil}) preserves space-time inversion symmetry $I_{\textrm{ST}}=C_{2z}T$ regardless of the presence of $\lambda_R$ and $\lambda_m$, Eq.~(\ref{eq_supHamil}) can be brought to a real gauge by unitary transformation $U_{\textrm{real}}$ satisfying
\begin{equation}
U_{\textrm{real}}(C_{2z}T)U_{\textrm{real}}^{\dagger}=K~,
\label{eq_realIst}
\end{equation}
where $K$ is complex conjugation.
Here, $C_{2z}$ and $T$ symmetry operators for Eq.~(\ref{eq_supHamil}) are
\begin{align}
C_{2z}=-i\sigma_z\otimes\mathbb{1}_3~,\qquad T=i\sigma_y\otimes\mathbb{1}_3K~,
\label{eq_c2zt}
\end{align}
where $\bm{\sigma}=(\sigma_x,\sigma_y,\sigma_z)$ are Pauli matrices acting on spin space and $\mathbb{1}_3$ is an identity matrix in sublattice space.
For these operators, $U_{\textrm{real}}$ satisfying Eq.~(\ref{eq_realIst}) is
\begin{equation}
\label{eq_Ureal}
U_{\textrm{real}}=\sqrt{\frac{1}{2}}
\begin{pmatrix}
1&i\\
i&1
\end{pmatrix}
\otimes\mathbbm{1}_3~.
\end{equation}
In this real gauge, the Hamiltonian is real, and its eigenstates can therefore be chosen real as well. However, choosing a real gauge is not enough to properly construct the Euler bands. We further need to match the orientation of the eigenstates consistently throughout the BZ~\cite{ahn2019failure}.
\begin{figure}[t]
\centering
\includegraphics[scale=0.5]{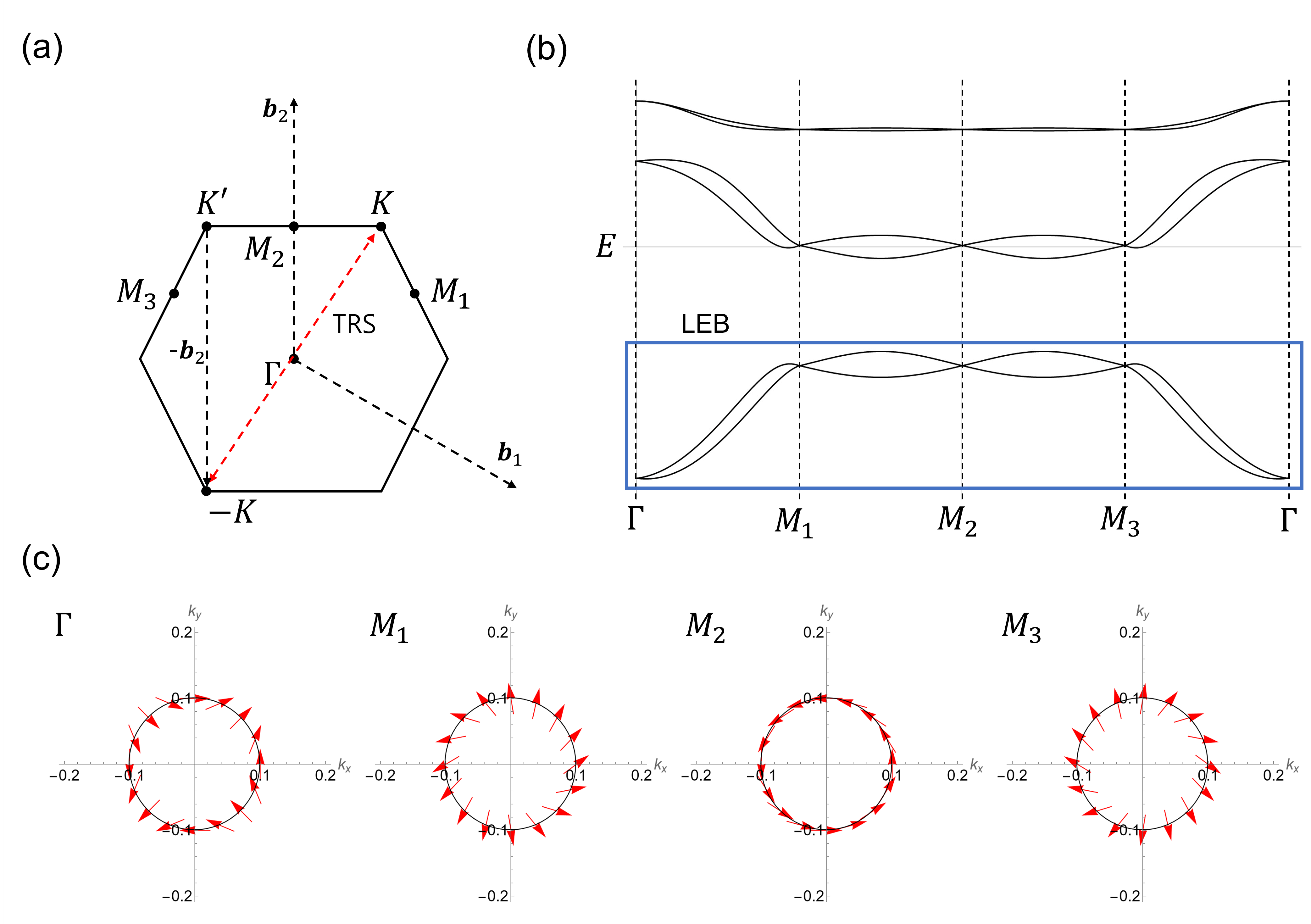}
\caption{{\bf The first Brillouin zone of spinful Kagome lattice, band structure and pseudospin texture of the LEB for $\lambda_R=0.1$ and $\lambda_m=0$.} (a) Position of the high symmetry momenta $\Gamma$, $M_1$, $M_2$, and $M_3$ points and reciprocal lattice vectors are shown. Here, the reciprocal lattice vectors are $\bm{b}_1=\frac{4\pi}{\sqrt{3}a}(\frac{\sqrt{3}}{2},-\frac{1}{2})$ and $\bm{b}_2=\frac{4\pi}{\sqrt{3}a}(0,1)$ with lattice constant $a$. (b) Band structure for $a=1,~t=1,~\lambda_{soc}=0.15,~\textrm{and}~\lambda_R=0.1$. Dirac crossings are placed at the $\Gamma$, $M_1$, $M_2$, and $M_3$ points.}
\label{fig_BZrashbaBS}
\end{figure}

Orientation matching of the real vector bundle is implemented as follows. We first consider the line connecting two momenta $\bm{p}$ and $\bm{q}$, and choose intermediate momenta $\bm{k}_i~(i=1,2,\dots,N)$ such that $\bm{p}<\bm{k}_1<\bm{k}_2<\dots<\bm{k}_N<\bm{q}$. At each momentum, we compute the two real eigenstates that belong to the Euler bundle. For each pair of neighboring momenta, we construct the $2\times2$ overlap matrix between the corresponding eigenstate doublets, for example $\{\ket{u(\bm{p})}_1,~\ket{u(\bm{p})}_2\}$ and $\{\ket{u(\bm{k}_1)}_1,~\ket{u(\bm{k}_1)}_2\}$. If the determinant of this matrix is $+1$, the orientation is already consistent and the order of the states is kept. If the determinant is $-1$, the order of the two states at the second momentum is exchanged. Repeating this procedure along the path yields an orientation-consistent basis connecting $\bm{p}$ and $\bm{q}$. This step is essential in Euler band systems and removes the residual gauge ambiguity associated with the ordering of the real eigenstates.

\section{Band structure of the Euler model Hamiltonian with Rashba coupling}
\label{supsec_BandStr}

In the Rashba-only case, $\lambda_R\neq0$ and $\lambda_m=0$, the system preserves $C_{2z}$ and $T$ symmetries, while inversion symmetry $P$ is broken by Rashba spin-orbit coupling. In this case, four Dirac crossings appear at the $\Gamma$, $M_1$, $M_2$, and $M_3$ points of the Brillouin zone, as shown in Figure~\ref{fig_BZrashbaBS}(a), and the corresponding band structure is shown in Figure~\ref{fig_BZrashbaBS}(b). To construct the effective Hamiltonian of the lower Euler bands (LEB), we follow the procedure described in Sec.~\ref{supsec_OrMatch} to fix the orientation of the real vector bundle.

After fixing the orientation, we project Eq.~(\ref{eq_supHamil}) onto the two-lowest-band subspace at each node. For instance, the projected Hamiltonian at $\Gamma$ is given by
\begin{equation}
\label{eq_realHgam}
H_{\Gamma,\textrm{real}}=
\begin{pmatrix}
\leftindex_1{\bra{u_\Gamma}}H_{\textrm{real}}\ket{u_\Gamma}_1&\leftindex_1{\bra{u_\Gamma}}H_{\textrm{real}}\ket{u_\Gamma}_2\\
\leftindex_2{\bra{u_\Gamma}}H_{\textrm{real}}\ket{u_\Gamma}_1&\leftindex_2{\bra{u_\Gamma}}H_{\textrm{real}}\ket{u_\Gamma}_2
\end{pmatrix}~,
\end{equation}
where $H_{\textrm{real}}=U_{\textrm{real}}HU_{\textrm{real}}^{\dagger}$ and $\ket{u_{\Gamma}}_i~(i=1,2)$ denotes an oriented orthonormal basis for the twofold-degenerate eigenspace at the $\Gamma$ point. Expanding the projected Hamiltonians to linear order in $\bm{k}$ around $\Gamma$, $M_1$, $M_2$, and $M_3$, we obtain four effective $2\times2$ Dirac Hamiltonians. From these effective low-energy Hamiltonians, we extract the pseudospin textures shown in Figure~\ref{fig_BZrashbaBS}(c), where the pseudospin vector is defined as
\begin{align}
\frac{1}{2}\left(\textrm{Tr}\left[H_{\textrm{eff}}\rho_x\right],\textrm{Tr}\left[H_{\textrm{eff}}\rho_y\right]\right)
\end{align}
around each node. Here, $H_{\textrm{eff}}\in\{H_{\Gamma,\textrm{real}},H_{M1,\textrm{real}},H_{M2,\textrm{real}},H_{M3,\textrm{real}}\}$ and $\bm{\rho}=(\rho_x,\rho_y,\rho_z)$ are Pauli matrices acting in the pseudospin space describing the two bands crossing at each valley. Figure~\ref{fig_BZrashbaBS}(c) shows that the $\Gamma$ point has vorticity $-1$, whereas $M_1$, $M_2$, and $M_3$ points have vorticity $+1$. The total vorticity is therefore +2, corresponding to $e_2=1$ for the LEB.

\section{Analysis of the Euler model Hamiltonian with in-plane magnetism}
\label{subsec_BEuler}
\begin{figure}[b]
\centering
\includegraphics[width=\linewidth]{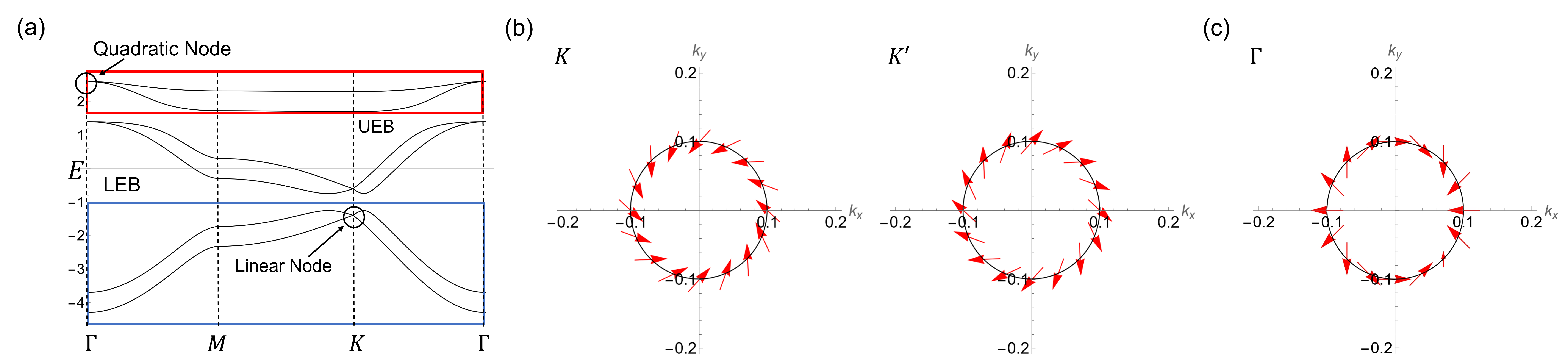}
\caption{{\bf Band structure and pseudospin textures of Euler band model for \bm{$\lambda_R=0,~\lambda_m=0.3$} at the $K$, $K'$ and $\Gamma$ points.} (a) The band structure. LEB (blue box) possesses two Dirac nodes at the $K$ and $K'$ point with the same vorticity $+1$, respectively. UEB (red box) posses a single quadratic node at the $\Gamma$ point with the vorticity $+2$. (b) Vorticity configurations at the $K$ and $K'$ points. Both points have the vorticity $+1$, yielding total vorticity $+2$ and $e_2=1$. (c) Vorticity configuration at the $\Gamma$ point. At the $\Gamma$ point, pseudospin rotates two times, implying vorticity +2.}
\label{fig_magneticps}
\end{figure}

In this section, we consider the case $\lambda_R=0$ and $\lambda_m\neq0$, whose band structure is shown in Figure~\ref{fig_spinfulKagome}(b). As in Sec.~\ref{supsec_BandStr}, we first fix the orientation of the eigenstates at the $K$ and $K'$ points where the Dirac nodes lie on the LEB and then project Eq.~(\ref{eq_supHamil}) onto the corresponding two-band subspaces, obtaining the projected real Hamiltonians at $K$ and $K'$ points,
\begin{equation}
\label{eq_HamKreal}
H_{K,\textrm{real}}=
\begin{pmatrix}
\leftindex_1{\bra{u_K}}H_{\textrm{real}}\ket{u_K}_1&\leftindex_1{\bra{u_K}}H_{\textrm{real}}\ket{u_K}_2\\
\leftindex_2{\bra{u_K}}H_{\textrm{real}}\ket{u_K}_1&\leftindex_2{\bra{u_K}}H_{\textrm{real}}\ket{u_K}_2
\end{pmatrix},~
H_{K',\textrm{real}}=
\begin{pmatrix}
\leftindex_1{\bra{u_{K'}}}H_{\textrm{real}}\ket{u_{K'}}_1&\leftindex_1{\bra{u_{K'}}}H_{\textrm{real}}\ket{u_{K'}}_2\\
\leftindex_2{\bra{u_{K'}}}H_{\textrm{real}}\ket{u_{K'}}_1&\leftindex_2{\bra{u_{K'}}}H_{\textrm{real}}\ket{u_{K'}}_2
\end{pmatrix}~.
\end{equation}
The resulting $2\times2$ Hamiltonians show that the two Dirac nodes at $K$ and $K'$ points both carry the same vorticity, as illustrated in Figure~\ref{fig_magneticps}(b).

To express the Dirac Hamiltonians, Eq.~(\ref{eq_HamKreal}), in the conventional $\rho_x$, $\rho_y$ basis, we apply the unitary transformation
\begin{equation}
\label{eq_Urot}
U_{\textrm{rot}}=\frac{1}{\sqrt{2}}
\begin{pmatrix}
e^{i \pi/4}&e^{-i \pi/4}\\
e^{i \pi/4}&-e^{-i \pi/4}
\end{pmatrix}~.
\end{equation}
This transformation converts the two real Pauli matrices appearing the projected real Hamiltonian Eq.~(\ref{eq_HamKreal}) into $\rho_x$ and $\rho_y$, so that the Hamiltonians have the usual two-component Dirac form. After this transformation, the coefficients of $\rho_x$ and $\rho_y$ can be linear combinations of the momentum components around each valley. We then choose local momentum coordinates near each Dirac point $K$ and $K'$ such that these two linear combinations are denoted by $k_x$ and $k_y$. This step is simply a rotation of the local momentum axes and does not change the physical winding of the pseudospin texture. In these local coordinates, the two valley Hamiltonians become
\begin{align}
H_K=v\left(k_x\rho_x+k_y\rho_y\right)~,\quad H_K'=-v\left(k_x\rho_x+k_y\rho_y\right)~.
\end{align}
Combining these Hamiltonians, we obtain
\begin{align}
H_{\textrm{LEB}}&\equiv
\begin{pmatrix}
H_K&0\\
0&H_K'
\end{pmatrix}\nonumber
\\
&=v
\begin{pmatrix}
k_x\rho_x+k_y\rho_y&0\\
0&-k_x\rho_x-k_y\rho_y
\end{pmatrix}\nonumber
\\
&=v\Pi_z\otimes\bm{k}\cdot\bm{\rho}
\label{eq_4x4Ham}
\end{align}
in the oriented basis $\{\ket{u_K}_1,\ket{u_K}_2,\ket{u_{K'}}_1,\ket{u_{K'}}_2\}$ of LEB, where
\begin{equation}
v=\sqrt{\frac{3t^2\lambda_m^2}{4(\lambda_m^2+3\lambda_{soc}^2)}}~.
\label{eq_vel}
\end{equation}
Thus, the LEB is described by two Dirac cones with identical vorticity.

In the presence of the in-plane magnetism, $C_{2z}$ and $T$ symmetries are broken separately, while $P$ symmetry and the combined symmetry $C_{2z}T$ remain preserved. Therefore, to analyze the effective $4\times4$ Hamiltonian in Eq.~(\ref{eq_4x4Ham}), we investigate the projected forms of $P$ and $C_{2z}T$ symmetries. In the full $6\times6$ spinful kagome model, these operators are given by
\begin{equation}
\label{eq_symOp}
P=\mathbbm{1}_6~,\quad C_{2z}T=-i\sigma_x\otimes\mathbbm{1}_3K~.
\end{equation}
Since inversion maps $K$ to $K'$ up to the reciprocal lattice vector $-\bm{b}_2$ [see Figure~\ref{fig_BZrashbaBS}(a)], we have
\begin{equation}
\label{eq_invTrf}
PH(K)P^{-1}=H(-K)=H(K'-\bm{b}_2)=V^{-1}(-\bm{b}_2)H(K')V(-\bm{b}_2)~,
\end{equation}
or equivalently
\begin{equation}
\label{eq_KpHam}
H(K')=[V(-\bm{b}_2)P]H(K)[V(-\bm{b}_2)P]^{-1}
\end{equation}
with
\begin{equation}
\label{eq_Vmat}
V(\bm{k})\equiv e^{i\bm{k}\cdot\bm{\tau}_{\alpha}}\delta_{\alpha\beta}~,
\end{equation}
where $\bm{\tau}_{\alpha}$ is the position of sublattice $\alpha$ in unit cell, $\alpha\in\{A\uparrow,B\uparrow,C\uparrow,A\downarrow,B\downarrow,C\downarrow\}$.
Therefore, the inversion symmetry operator relating the two valleys $K$ and $K'$ is
\begin{equation}
\label{eq_conInvop}
P^{KK'}=V(-\bm{b}_2)P~.
\end{equation}
Transforming this operator to the real gauge gives
\begin{equation}
\label{eq_realconInvop}
P_{\textrm{real}}^{KK'}=U_{\textrm{real}}P^{KK'}U_{\textrm{real}}^{\dagger}~.
\end{equation}

We then project $P_{\textrm{real}}^{KK'}$ onto the oriented LEB basis
\begin{align}
\{\ket{u_K}_1,\ket{u_K}_2,\ket{u_{K'}}_1,\ket{u_{K'}}_2\}
\label{eq_orbasis}
\end{align}
to obtain the inversion symmetry operator connecting $K$ and $K'$ points in LEB subspace.
Because $P_{\textrm{real}}^{KK'}$ exchanges the two valleys, its matrix elements are determined by how it maps the basis states between the $K$ and $K'$ sectors:
\begin{align}
\label{eq_KKpconInv}
&P_{\textrm{LEB,real}}\ket{u_{K_1}}=c_1\ket{u_{K_1'}}+c_2\ket{u_{K_2'}}\nonumber
\\
&P_{\textrm{LEB,real}}\ket{u_{K_2}}=c_3\ket{u_{K_1'}}+c_4\ket{u_{K_2'}}\nonumber
\\
&P_{\textrm{LEB,real}}\ket{u_{K_1'}}=c_5\ket{u_{K_1}}+c_6\ket{u_{K_2}}\nonumber
\\
&P_{\textrm{LEB,real}}\ket{u_{K_2'}}=c_7\ket{u_{K_1}}+c_8\ket{u_{K_2}}~.
\end{align}
This gives
\begin{equation}
P_{\textrm{LEB,real}}=
\begin{pmatrix}
\label{eq_realPmat}
0&0&c_5&c_7\\
0&0&c_6&c_8\\
c_1&c_3&0&0\\
c_2&c_4&0&0
\end{pmatrix}~,
\end{equation}
and solving Eq.~(\ref{eq_KKpconInv}) yields
\begin{equation}
\label{eq_realPmatnum}
P_{\textrm{LEB,real}}=
\begin{pmatrix}
0&0&-1&0\\
0&0&0&-1\\
-1&0&0&0\\
0&-1&0&0
\end{pmatrix}~.
\end{equation}
To express the projected inversion operator in the same convention as the low-energy Hamiltonian in Eq.~(\ref{eq_4x4Ham}), we now apply the pseudospin-basis rotation $U_{\textrm{rot}}$ to each valley sector. In addition, for later comparison with graphene and for the diagrammatic calculation, we exchange the order of the two basis states in the $K'$ valley. Namely, we use the basis $\{\ket{u_{K_1}},\ket{u_{K_2}},\ket{u_{K_2'}},\ket{u_{K_1'}}\}$ instead of $\{\ket{u_{K_1}},\ket{u_{K_2}},\ket{u_{K_1'}},\ket{u_{K_2'}}\}$. This exchange is only a relabeling of the pseudospin basis in the second valley and does not change any physical observable. In this basis, the projected inversion symmetry of LEB is written as
\begin{equation}
\label{eq_LEBInv}
P_{\textrm{LEB}}=-\Pi_x\otimes\rho_x~.
\end{equation}

The projection of $C_{2z}T$ symmetry is simpler because $C_{2z}T$ symmetry acts within a single valley without exchanging $K$ and $K'$:
\begin{equation}
\label{eq_C2ztrf}
(C_{2z}T)H(K)(C_{2z}T)^{-1}=H(K)~.
\end{equation}
Therefore, in the projected LEB subspace,
\begin{equation}
\label{eq_C2zOp}
(C_{2z}T)_{\textrm{LEB}}=i\mathbb{1}_2\otimes\rho_xK~,
\end{equation}
and the projection of the \textit{effective} time-reversal symmetry onto the LEB becomes
\begin{equation}
\label{eq_effTLEB}
T_{\textrm{LEB}}^*=P_{\textrm{LEB}}(C_{2z}T)_{\textrm{LEB}}=-i\Pi_x\otimes\mathbbm{1}_2K
\end{equation}
in $\{\ket{u_{K_1}},\ket{u_{K_2}},\ket{u_{K_2'}},\ket{u_{K_1'}}\}$ basis.
In this way, the relevant symmetry operators projected onto the LEB are obtained.

It is also useful to identify the valley-local time-reversal symmetry $T_{\textrm{SV}}$ of a single Dirac cone. The single-valley Hamiltonian $k_x\rho_x+k_y\rho_y$ admits the local antiunitary symmetry $i\rho_yK$. Since the projected Hamiltonians at $K$ and $K'$ differ only by the overall sign, the same form of TRS applies to both valleys, which gives
\begin{equation}
\label{eq_TSV}
T_{\textrm{SV}}=i\mathbbm{1}_2\otimes\rho_yK~.
\end{equation}

Finally, because the full two-valley system preserves $T_{\textrm{LEB}}^*$, any disorder potential included in the projected theory must satisfy
\begin{align}
T_{\textrm{LEB}}^*V_{\textrm{dis}}(T_{\textrm{LEB}}^*)^{-1}=V_{\textrm{dis}}~.
\label{eq_dissym}
\end{align}
Among the $16$ possible disorder potentials, only $10$ of them satisfy Eq.~(\ref{eq_dissym}):
\begin{align}
\label{tempeq_DisOp}
&V_{\textrm{sc}}=u_{00}\mathbbm{1}_4~,\nonumber\\
&V_{\textrm{intra}}=u_{xz}\Sigma_x\Lambda_z,~u_{y0}\Sigma_y\Lambda_0,~u_{z0}\Sigma_z\Lambda_0~,\nonumber\\
&V_{\textrm{inter}}=u_{0x}\Sigma_0\Lambda_x,~u_{0y}\Sigma_0\Lambda_y,~u_{yx}\Sigma_y\Lambda_x,~u_{yy}\Sigma_y\Lambda_y,~u_{zx}\Sigma_z\Lambda_x,~u_{zy}\Sigma_z\Lambda_y~.
\end{align}
Furthermore, these disorder potentials should satisfy $P_{\textrm{LEB}}$ and $(C_{2z}T)_{\textrm{LEB}}$ of the magnetic Euler bands. Therefore, disorder potentials that satisfy all these symmetries are
\begin{align}
\label{eq_DisOp}
&V_{\textrm{sc}}=u_{00}\mathbbm{1}_4~,\nonumber\\
&V_{\textrm{intra}}=u_{xz}\Sigma_x\Lambda_z,~u_{y0}\Sigma_y\Lambda_0~,\nonumber\\
&V_{\textrm{inter}}=~u_{0y}\Sigma_0\Lambda_y,~u_{yy}\Sigma_y\Lambda_y,~u_{zx}\Sigma_z\Lambda_x~.
\end{align}
These are the disorder potentials used in the diagrammatic calculation of the quantum correction to the conductivity of LEB having $\lambda_R=0$ and $\lambda_m\neq0$.

For completeness, we also study the vorticity of the upper Euler bands (UEB), two uppermost bands of Eq.~(\ref{eq_supHamil}) with $\lambda_R=0$ and $\lambda_m\neq0$ using the similar projection process. Projecting Eq.~(\ref{eq_supHamil}) onto the $\Gamma$ point using the two uppermost bands yields
\begin{align}
\label{eq_lowEHamUEB}
H_{\textrm{UEB}}&=U_{\textrm{rot}}
\begin{pmatrix}
\leftindex_1{\bra{u_{\Gamma}}}H_\textrm{real}\ket{u_{\Gamma}}_1&\leftindex_1{\bra{u_{\Gamma}}}H_\textrm{real}\ket{u_{\Gamma}}_2\\
\leftindex_2{\bra{u_{\Gamma}}}H_\textrm{real}\ket{u_{\Gamma}}_1&\leftindex_2{\bra{u_{\Gamma}}}H_\textrm{real}\ket{u_{\Gamma}}_2
\end{pmatrix}
U_{\textrm{rot}}^{\dagger}\nonumber
\\
&=-\mu\left[(k_x^2-k_y^2)\rho_x+2k_x k_y\rho_y\right]~.
\end{align}
The corresponding pseudospin texture, shown in Figure~\ref{fig_magneticps}(c ), has vorticity $+2$, as expected.
\begin{figure}[b]
\centering
\includegraphics[scale=0.5]{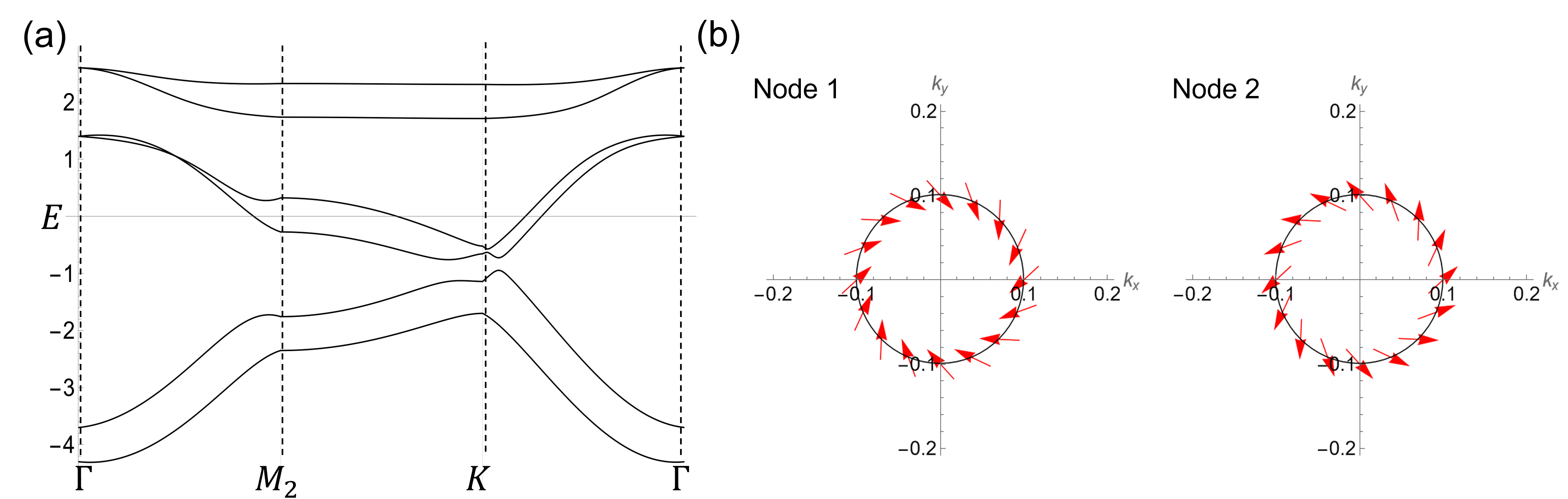}
\caption{{\bf Band structure and pseudospin textures of Euler band model for \bm{$\lambda_R=0.1,~\lambda_m=0.3$} at node 1 and node 2.} (a) The band structure. Nodes are not placed at the $K$ and $K'$ points anymore, due to the Rashba term and in-plane magnetism that breaks both \textit{physical} and \textit{effective} TRS. (b) Pseudospin textures of the nodes shifted away from $K$ and $K'$ points. Each point has vorticity of +1, yielding total vorticity +2 and $e_2=1$ as before.}
\label{fig_bspspinboth}
\end{figure}

\section{Analysis of the Euler model Hamiltonian with both Rashba coupling and an in-plane magnetism}
\label{supsec_noTRS}

We now consider the case $\lambda_R\neq0$ and $\lambda_m\neq0$, in which both Rashba coupling and the in-plane magnetism are present. In this regime, both \textit{physical} TRS and the \textit{effective} TRS are broken. As a result, the Dirac nodes are no longer constrained to high-symmetry points and can in general appear at generic momenta in the BZ, as illustrated in Figure~\ref{fig_anomaly}(d).

For representative parameters $\lambda_R=0.1$ and $\lambda_m=0.3$, we numerically find two Dirac nodes located at
\begin{align}
(2.15443,-3.07544)a^{-1}~,\quad (-2.15443,-3.07544)a^{-1}~.
\end{align}
The corresponding band structure and pseudospin textures are shown in Figure~\ref{fig_bspspinboth}. As seen in Figure~\ref{fig_bspspinboth}(b), each node carries vorticity $+1$. Therefore, although the node positions are shifted away from $K$ and $K'$, the total vorticity remains $+2$, and the Euler number is unchanged at $e_2=1$.

\section{Symmetry analysis of the localization correction}
\label{supsec_symAnal}
Before presenting the explicit Cooperon calculation, it is useful to classify the localization correction by the antiunitary symmetries. When only scalar disorder potential (disorder potential proportional to the identity matrix in the pseudospin basis) is present, scattering is confined within each valley. The relevant antiunitary symmetry is then the single-valley time-reversal operation $T_{\textrm{SV}}$ introduced in Eq.~(\ref{eq_TSV}). Because $T_{\textrm{SV}}^2=-1$, the system belongs to the symplectic class and exhibits WAL, independent of $\lambda_R$ and $\lambda_m$. When general intravalley disorder potentials that cause the scattering between different pseudospin components in a single valley are added, $T_{\textrm{SV}}$ is broken. Since no other antiunitary symmetry acts within an isolated valley, the system now belongs to the unitary class, and the singular localization correction disappears.

Once intervalley disorder potentials are introduced, the valleys are coupled and the relevant symmetry is that of the full Hamiltonian. For $\lambda_R\neq0$  and $\lambda_m=0$, \textit{physical} TRS $T$ is preserved, with $T^2=-1$, so the system remains in the symplectic class and continues to show WAL. For $\lambda_R\neq0$ and $\lambda_m\neq0$, both \textit{physical} and \textit{effective} TRS are absent. Intervalley scattering therefore does not restore any antiunitary symmetry constraint, and the system remains in the unitary class, with no singular localization correction. By contrast, for $\lambda_R=0$ and $\lambda_m\neq0$, the combination of $C_{2z}T$ and inversion symmetries restores the \textit{effective} TRS $T^*=M_zT$. Once intervalley scattering couples the two nodes, this becomes the relevant antiunitary symmetry of the two-valley system. Because $(T^*)^2=+1$, the system belongs to the orthogonal class and exhibits WL correction. These symmetry-based results are summarized in Table~\ref{tb_2}.
\begin{table}[h]
\caption{{\bf Classification of the spinful Euler band models.} The localization class is determined by the antiunitary symmetry relevant to each disorder channel. For instance, spinful system with both $C_{2z}$ and $T$ symmetries (the third row of the table) belongs to the symplectic class if the applied disorder potentials are scalar+intervalley disorder potentials. On the other hand, spinful system with both $C_{2z}T$ and $P$ symmetries (the fifth row of the table) belongs to the orthogonal class when the scalar+intervalley disorder potentials are applied. For each row, we retain only the disorder potentials compatible with the symmetries listed in that row ($C_{2z}$ and $T$ for the third row, $C_{2z}T$ for the fourth row, and $C_{2z}T$ and $P$ for the fifth row)}
\begin{tabular}{c|c|c|c|c}
\hline\hline
\multicolumn{2}{c|}{}&
\multicolumn{3}{c}{Disorder}
\\ \hline
System&
Symmetry &
Scalar&
Scalar+Intravalley&
Scalar+Intervalley
\\ \hline
\multirow{3}{*}{Spinful}&
$C_{2z},T$&
Symplectic&
Unitary&
Symplectic
\\
&
$C_{2z}T$&
Symplectic&
Unitary&
Unitary
\\
&
$C_{2z}T,P$&
Symplectic&
Unitary&
Orthogonal
\\
\hline\hline
\end{tabular}
\label{tb_2}
\end{table}

\section{Cooperon correction diagram Analysis}
\label{subsec_coopcalc}
Throughout Sec.~\ref{subsec_coopcalc}--\ref{supsec_dg}, we work in the static retarded-advanced formulation appropriate for the DC conductivity. We use the retarded-advanced expression obtained by analytically continuing the Kubo formula and subsequently taking the limit $\Omega\rightarrow0$, where $\Omega$ denotes the external frequency of the current-current response. Accordingly, all disorder-averaged Green's functions, current vertices, and Cooperons below are evaluated at zero external frequency.
To evaluate the disorder-induced correction to the DC conductivity in the weak disorder regime, we use the LEB Hamiltonian of $\lambda_R=0$ and $\lambda_m\neq0$ in Eq.~(\ref{eq_4x4Ham}), which preserves the \textit{effective} TRS in Eq.~(\ref{eq_effTLEB}), as our representative example. The same procedure can be generalized straightforwardly to the case $\lambda_R\neq0$ and $\lambda_m\neq0$, where both \textit{physical} and \textit{effective} TRS are broken. We label the LEB basis as
\begin{align}
\label{eq_abbreviationLEB}
\{1,2,3,4\}\equiv\{\ket{u_K}_1,\ket{u_K}_2,\ket{u_{K'}}_2,\ket{u_{K'}}_1\}~.
\end{align}
This implies that the order of the two basis states in the $K'$ valley is exchanged relative to the oriented basis of Eq.~(\ref{eq_orbasis}), as discussed in Sec.~\ref{subsec_BEuler}. In this convention, graphene is described by
\begin{align}
\label{eq_swapGraph}
H_{\textrm{graphene}}=v\Pi_z\otimes\bm{k}\cdot\bm{\rho}~,
\end{align}
while the LEB Hamiltonian becomes
\begin{align}
\label{eq_swapLEB}
\widetilde{H}_{\textrm{LEB}}\equiv v\left[\frac{1}{2}(\mathbbm{1}_2+\Pi_z)\otimes\bm{k}\cdot\bm{\rho}-\frac{1}{2}(\mathbbm{1}_2-\Pi_z)\otimes(\bm{k}\cdot\bm{\rho})^T\right]~.
\end{align}
This representation keeps the disorder potentials in the same form as in the previous study of graphene~\cite{mccann2006weak}, while encoding the different relative vorticity through the valley structure of the Hamiltonian.

Before analyzing the Cooperon contribution, we first evaluate the ladder-type vertex correction shown in Figure~\ref{fig_manybody}(b). For the transport along the $j$-direction, the renormalized velocity operator $\tilde{v}_j$ satisfies
\begin{align}
(\tilde{v}_j)_{\alpha'\alpha}=(v_j)_{\alpha'\alpha}+\sum_{\bm{k}}\sum_{\beta,\gamma,\beta',\gamma'}V_{\beta\alpha}V_{\alpha'\beta'}\left[G_R(\bm{k})\right]_{\gamma\beta}\left[G_A(\bm{k})\right]_{\beta'\gamma'}(\tilde{v}_j)_{\gamma'\gamma}
\label{eq_VCEq}
\end{align}
with
\begin{align}
G_{R/A}(\bm{k})&\equiv\frac{1}{\epsilon_{R/A}-\widetilde{H}_{\textrm{LEB}}}~\quad \left(\epsilon_R\equiv E_F+\frac{i}{2\tau_{00}}~,\quad\epsilon_A\equiv E_F-\frac{i}{2\tau_{00}}\right)\nonumber
\\
&=\frac{\epsilon_{R/A}+\widetilde{H}_{\textrm{LEB}}}{\epsilon_{R/A}^2-v^2k^2}\nonumber
\\
&=\frac{1}{\epsilon_{R/A}^2-v^2k^2}\begin{pmatrix}
\epsilon_{R/A}&v(k_x-ik_y)&0&0\\
v(k_x+ik_y)&\epsilon_{R/A}&0&0\\
0&0&\epsilon_{R/A}&-v(k_x+ik_y)\\
0&0&-v(k_x-ik_y)&\epsilon_{R/A}
\end{pmatrix}
\label{eq_GAV}
\end{align}
and
\begin{align}
v_j\equiv\frac{\partial \widetilde{H}_{\textrm{LEB}}}{\partial k_j}~.
\end{align}
Assuming that the scalar disorder potential $V_{\textrm{sc}}=u_{00}\Sigma_0\Lambda_0$ dominates the system, Eq.~(\ref{eq_VCEq}) reduces to
\begin{align}
(\tilde{v}_j)_{\alpha'\alpha}&=(v_j)_{\alpha'\alpha}+u_{00}^2\sum_{\bm{k}}\sum_{\gamma,\gamma'}\left[G_R(\bm{k})\right]_{\gamma\alpha}\left[G_A(\bm{k})\right]_{\alpha'\gamma'}(\tilde{v}_j)_{\gamma'\gamma}\nonumber
\\
&\equiv(v_j)_{\alpha'\alpha}+\sum_{\gamma,\gamma'}Z_{\gamma\alpha\alpha'\gamma'}(\tilde{v}_j)_{\gamma'\gamma}~,
\end{align}
where
\begin{align}
Z_{\alpha\beta\alpha'\beta'}\equiv u_{00}^2\sum_{\bm{k}}\left[G_R(\bm{k})\right]_{\alpha\beta}\left[G_A(\bm{k})\right]_{\alpha'\beta'}~.
\end{align}
The momentum integral can be evaluated by the contour integral method. A representative example is the $(1111)$ component of $Z$, whose explicit evaluation is as follows,
\begin{align}
Z_{1111}&=u_{00}^2\sum_{\bm{k}}\left[G_R(\bm{k})\right]_{11}\left[G_A(\bm{k})\right]_{11}=\sum_{\bm{k}}\frac{\epsilon_R\epsilon_Au_{00}^2}{(\epsilon_R^2-v^2k^2)(\epsilon_A^2-v^2k^2)}\nonumber
\\
&=\int_0^{\infty}\frac{kdk}{2\pi}\int_0^{2\pi}\frac{d\theta}{2\pi}\frac{\epsilon_R\epsilon_Au_{00}^2}{(\epsilon_R^2-v^2k^2)(\epsilon_A^2-v^2k^2)}~,\quad \epsilon=vk\nonumber
\\
&=\int_0^{\infty}d\epsilon\frac{\epsilon}{2\pi v^2}\frac{\epsilon_R\epsilon_Au_{00}^2}{(\epsilon^2-\epsilon_R^2)(\epsilon^2-\epsilon_A^2)}\nonumber
\\
&\equiv\int_0^{\infty}d\epsilon\rho(\epsilon)\frac{\epsilon_R\epsilon_Au_{00}^2}{(\epsilon^2-\epsilon_R^2)(\epsilon^2-\epsilon_A^2)}~\quad \left(\rho(\epsilon)\equiv\frac{\epsilon}{2\pi v^2}\right)\nonumber
\\
&\simeq N_0\int_0^{\infty}d\epsilon\frac{E_F^2u_{00}^2}{(2E_F)^2\left(\epsilon-E_F-\frac{i}{2\tau_{00}}\right)\left(\epsilon-E_F+\frac{i}{2\tau_{00}}\right)}~\quad\left(N_0\equiv\rho(E_F)=\frac{|E_F|}{2\pi v^2}\right)\nonumber
\\
&=\frac{N_0}{4}\int_{-E_F}^{\infty}d\epsilon\frac{u_{00}^2}{\left(\epsilon-\frac{i}{2\tau_{00}}\right)\left(\epsilon+\frac{i}{2\tau_{00}}\right)}\nonumber
\\
&=\frac{N_0u_{00}^2}{4}2\pi i\frac{1}{\frac{i}{\tau_{00}}}\nonumber
\\
&=\frac{1}{2}~,
\label{eq_zex}
\end{align}
where $\tau_{00}^{-1}=\pi N_0u_{00}^2$ is the elastic scattering rate due to the scalar disorder potential, and $N_0$ is the density of states at the Fermi energy $E_F$. We can approximate $(\epsilon+\epsilon_R)(\epsilon+\epsilon_A)\simeq4E_F^2$ since the only valid integration region is $\epsilon=E_F+\delta$ with $|\delta|\le (\textrm{few}~\frac{1}{2\tau_{00}})$. Concretely,
\begin{align}
(\epsilon+\epsilon_R)(\epsilon+\epsilon_A)&=\left(2E_F+\delta+\frac{i}{2\tau_{00}}\right)\left(2E_F+\delta-\frac{i}{2\tau_{00}}\right)\nonumber
\\
&=(2E_F+\delta)^2+\frac{1}{4\tau_{00}^2}\nonumber
\\
&=4E_F^2\left[1+\mathcal{O}\left(\frac{1}{E_F\tau_{00}}\right)\right]\nonumber
\\
&\simeq4E_F^2~
\end{align}
in the weak disorder limit $E_F\tau_{00}\gg1$. Also, the density of states can be replaced by the constant $N_0$ within the energy window of width $1/\tau_{00}$ around the Fermi level, where the Green's functions are sharply peaked. In the final step of Eq.~(\ref{eq_zex}), we have also extended the lower limit of the energy integration from $-E_F$ to $-\infty$, where $\epsilon$ now denotes the deviation from the Fermi level. After the approximations above, the integrand is a product of retarded and advanced propagators sharply peaked at $\epsilon=0$ with width $1/2\tau_{00}$, decaying as $\epsilon^{-2}$ for $|\epsilon|\gg1/\tau_{00}$. The added region $\epsilon<-E_F$ thus contains no pole and carries no $\tau$ enhancement. Its contribution is of order $1/E_F$, to be compared with $2\pi\tau_{00}$ from the peak, so that the extension introduces a relative error of order $(E_F\tau_{00})^{-1}$, consistent with the weak-disorder limit $E_F\tau_{00}\gg1$. Once the limit is extended, the contour may be closed in either half-plane, and only a single pole contributes. The same approximation is used for similar calculations in the following sections. Evaluating the remaining components in the same way, we obtain the renormalized velocity
\begin{align}
\tilde{v}_j=2v_j
\end{align}
for $\widetilde{H}_{\textrm{LEB}}$.

We next determine which Cooperon channels contribute to the diagram in Figure~\ref{fig_manybody}(a). The conductivity correction takes the form
\begin{align}
\label{eq_dginXY}
\delta g_j=\frac{e^2}{\pi}\sum_{\{\alpha\}}\sum_{\bm{k}}\left[X_j(\bm{k})\right]_{\alpha\beta'}\left[Y_j(\bm{k})\right]_{\beta\alpha'}\sum_{\bm{q}}\left[C(\bm{q})\right]_{\alpha'\beta',\alpha\beta}~,
\end{align}
where
\begin{align}
\label{eq_xydefinsup}
&X_j(\bm{k})\equiv G_R(\bm{k})\tilde{v}_jG_A(\bm{k})~,\nonumber
\\
&Y_j(\bm{k})\equiv G_A(-\bm{k})\tilde{v}_jG_R(-\bm{k})~,
\end{align}
and
$\{\alpha\}\equiv\alpha,\beta,\alpha',\beta'$.

Because $\widetilde{H}_{\textrm{LEB}}$, the disorder-averaged Green's function, and the velocity operator are all valley diagonal,
\begin{align}
G_{R/A}=
\stackrel{\mbox{1~~2}~~~~~~\mbox{3~~4}}{\begin{pmatrix}
G_{R/A}^K&0\\
0&G_{R/A}^{K'}
\end{pmatrix}}~,\quad
v_j=
\stackrel{\mbox{1~2}~~~\mbox{3~4}}{\begin{pmatrix}
v_j^K&0\\
0&v_j^{K'}
\end{pmatrix}}~,
\end{align}
$X_j$ and $Y_j$ are also valley diagonal. As a result, matrix elements such as $\left(X_j\right)_{13}$ vanish, and only terms whose indices remain within the same valley block can contribute. Therefore, the number of possible Cooperons is reduced from $4^4$ to $4^3$.

A further reduction follows from angular averaging. We define the parity of the sublattice-like basis to be $+1$ for $\{1,4\}$ basis and $-1$ for $\{2,3\}$ basis. Then a weight factor involving an odd number of parity flips vanishes after angular integration. Here we consider $\left[X_x(\bm{k})\right]_{11}\left[Y_x(\bm{k})\right]_{12}$ as an example, but generalization to other matrix elements is straightforward. Analytic form of $\left[X_x(\bm{k})\right]_{11}\left[Y_x(\bm{k})\right]_{12}$ is
\begin{align}
\label{eq_exzeroXY}
\sum_{\bm{k}}\left[X_x(\bm{k}   )\right]_{11}\left[Y_x(\bm{k})\right]_{12}=\sum_{\bm{k}}\frac{4v^2\left[\epsilon_R(k_x+ik_y)+\epsilon_A(k_x-ik_y)\right]\left[\epsilon_R\epsilon_A+(k_x-ik_y)^2v^2\right]}{\left(\epsilon_R^2-v^2k^2\right)^2\left(\epsilon_A^2-v^2k^2\right)^2}~,
\end{align}
using Eq.~(\ref{eq_xydefinsup}). With $ke^{i\theta}\equiv k_x+ik_y$, the numerator of Eq.~(\ref{eq_exzeroXY}) becomes
\begin{align}
\label{eq_exnum}
4v^2k(e^{-i\theta}\epsilon_R\epsilon_A^2+e^{i\theta}\epsilon_R^2\epsilon_A+e^{-3i\theta}\epsilon_Av^2k^2+e^{-i\theta}\epsilon_Rv^2k^2)~,
\end{align}
which clearly shows the angular dependence of $\left[X_x(\bm{k})\right]_{11}\left[Y_x(\bm{k})\right]_{12}$. Therefore, the angular integration of Eq.~(\ref{eq_exzeroXY}) is zero. By contrast, for $\left[X_x(\bm{k})\right]_{11}\left[Y_x(\bm{k})\right]_{11}$, we compute
\begin{align}
\label{eq_exnonzeroXY}
\sum_{\bm{k}}\left[X_x(\bm{k})\right]_{11}\left[Y_x(\bm{k})\right]_{11}&=\sum_{\bm{k}}\frac{-4v^4k^2\left(\epsilon_Re^{-i\theta}+\epsilon_Ae^{i\theta}\right)\left(\epsilon_Ae^{-i\theta}+\epsilon_Re^{i\theta}\right)}{(\epsilon_R^2-v^2k^2)^2(\epsilon_A^2-v^2k^2)^2}\nonumber
\\
&=-4v^2\int\frac{d^2k}{4\pi^2}\frac{v^2k^2(\epsilon_R^2+\epsilon_A^2+2\epsilon_R\epsilon_A\cos2\theta)}{(\epsilon_R^2-v^2k^2)^2(\epsilon_A^2-v^2k^2)^2}\nonumber
\\
&=-4v^2\int_0^{\infty}\frac{kdk}{2\pi}\frac{v^2k^2(\epsilon_R^2+\epsilon_A^2)}{(\epsilon_R^2-v^2k^2)^2(\epsilon_A^2-v^2k^2)^2}\nonumber
\\
&=-4v^2\int_0^{\infty}d\epsilon\frac{\epsilon}{2\pi v^2}\frac{\epsilon^2(\epsilon_R^2+\epsilon_A^2)}{(\epsilon^2-\epsilon_R^2)^2(\epsilon^2-\epsilon_A^2)^2}\nonumber
\\
&\simeq-4v^2N_0\int_0^{\infty}d\epsilon\frac{\epsilon^2(\epsilon_R^2+\epsilon_A^2)}{(\epsilon+\epsilon_R)^2(\epsilon+\epsilon_A)^2(\epsilon-\epsilon_R)^2(\epsilon-\epsilon_A)^2}\nonumber
\\
&\simeq-4v^2N_0\int_0^{\infty}d\epsilon\frac{2E_F^4}{(2E_F)^2(2E_F)^2\left(\epsilon-E_F-\frac{i}{2\tau_{00}}\right)^2\left(\epsilon-E_F+\frac{i}{2\tau_{00}}\right)^2}\nonumber
\\
&=-\frac{v^2N_0}{2}\int_{-E_F}^{\infty}d\epsilon\frac{1}{\left(\epsilon-\frac{i}{2\tau_{00}}\right)^2\left(\epsilon+\frac{i}{2\tau_{00}}\right)^2}\nonumber
\\
&\simeq-\frac{v^2N_0}{2}\int_{-\infty}^{\infty}d\epsilon\frac{1}{\left(\epsilon-\frac{i}{2\tau_{00}}\right)^2\left(\epsilon+\frac{i}{2\tau_{00}}\right)^2}\nonumber
\\
&=-\frac{v^2N_0}{2}2\pi i\frac{d}{d\epsilon}\frac{1}{\left(\epsilon+\frac{i}{2\tau_{00}}\right)^2}\bigg|_{\epsilon=\frac{i}{2\tau_{00}}}\nonumber
\\
&=-\frac{v^2N_0}{2}2\pi i(-2i\tau_{00}^3)\nonumber
\\
&=-2\pi N_0v^2\tau_{00}^3~,
\end{align}
where we have used the same approximation as in Eq.~(\ref{eq_zex}). Extending this analysis to all matrix elements, we find that only parity-conserving weights remain. For $\alpha=1$, the nonzero terms are
\begin{align}
\label{eq_nonzeroweight}
&(X_j)_{11}(Y_j)_{11},~(X_j)_{11}(Y_j)_{22},~(X_j)_{12}(Y_j)_{12},~(X_j)_{12}(Y_j)_{21},\nonumber
\\
&(X_j)_{11}(Y_j)_{33},~(X_j)_{11}(Y_j)_{44},~(X_j)_{12}(Y_j)_{34},~(X_j)_{12}(Y_j)_{43}~.
\end{align}

Including the corresponding contributions for $\alpha=2,3,4$, we obtain $32$ nonvanishing Cooperon channels in total. For $j=x$, the conductivity correction Eq.~(\ref{eq_dginXY}) therefore reduces to the explicit expression as
\begin{align}
\label{eq_fullbubb}
\delta g_x=\frac{e^2}{\pi}\sum_{\bm{k},\bm{q}}&(X_{11}Y_{11}C_{11,11}+X_{11}Y_{22}C_{21,12}+X_{12}Y_{12}C_{22,11}+X_{12}Y_{21}C_{12,12}+X_{11}Y_{33}C_{31,13}\nonumber
\\
&+X_{11}Y_{44}C_{41,14}+X_{12}Y_{34}C_{42,13}+X_{12}Y_{43}C_{32,14}+X_{22}Y_{22}C_{22,22}+X_{22}Y_{11}C_{12,21}\nonumber
\\
&+X_{21}Y_{21}C_{11,22}+X_{21}Y_{12}C_{21,21}+X_{22}Y_{33}C_{32,23}+X_{22}Y_{44}C_{42,24}+X_{21}Y_{34}C_{41,23}\nonumber
\\
&+X_{21}Y_{43}C_{31,24}+X_{33}Y_{33}C_{33,33}+X_{33}Y_{44}C_{43,34}+X_{34}Y_{34}C_{44,33}+X_{34}Y_{43}C_{34,34}\nonumber
\\
&+X_{33}Y_{11}C_{13,31}+X_{33}Y_{22}C_{23,32}+X_{34}Y_{12}C_{24,31}+X_{34}Y_{21}C_{14,32}+X_{44}Y_{44}C_{44,44}\nonumber
\\
&+X_{44}Y_{33}C_{34,43}+X_{43}Y_{43}C_{33,44}+X_{43}Y_{34}C_{43,43}+X_{44}Y_{22}C_{24,42}+X_{44}Y_{11}C_{14,41}\nonumber
\\
&+X_{43}Y_{21}C_{13,42}+X_{43}Y_{12}C_{23,41})~,
\end{align}
where we set $X\equiv X_x$ and $Y\equiv Y_x$. Evaluating $\sum_{\bm{k}}X_{\alpha\beta'}Y_{\beta\alpha'}$ appearing in Eq.~(\ref{eq_fullbubb}) gives
\begin{align}
\label{eq_totweight}
\{&X_{11}Y_{11},X_{11}Y_{22},X_{11}Y_{33},X_{11}Y_{44},X_{12}Y_{34},X_{21}Y_{43},X_{22}Y_{11},X_{22}Y_{22},X_{22}Y_{33},X_{22}Y_{44},\nonumber
\\
&X_{33}Y_{11},X_{33}Y_{22},X_{33}Y_{33},X_{33}Y_{44},X_{34}Y_{12},X_{43}Y_{21},X_{44}Y_{11},X_{44}Y_{22},X_{44}Y_{33},X_{44}Y_{44}\}=-2\pi N_0v^2\tau_{00}^3~,\nonumber
\\
\{&X_{12}Y_{12},X_{21}Y_{21},X_{34}Y_{34},X_{43}Y_{43}\}=\pi N_0v^2\tau_{00}^3~,\nonumber
\\
\{&X_{12}Y_{21},X_{21}Y_{12},X_{34}Y_{43},X_{43}Y_{34}\}=2\pi N_0v^2\tau_{00}^3~,\nonumber
\\
\{&X_{12}Y_{43},X_{21}Y_{34},X_{34}Y_{21},X_{43}Y_{12}\}=-\pi N_0v^2\tau_{00}^3~,
\end{align}
where only the scalar disorder potential is assumed to be present. When additional channels are present, $\tau_{00}$ is replaced by the total scattering rate as described in the following section. Plugging Eq.~(\ref{eq_totweight}) to Eq.~(\ref{eq_fullbubb}), we get
\begin{align}
\delta g_x=e^2N_0v^2\tau_{00}^3\sum_{\bm{q}}(&-2C_{11,11}-2C_{21,12}+C_{22,11}+2C_{12,12}-2C_{31,13}-2C_{41,14}-2C_{42,13}-C_{32,14}\nonumber
\\
&-2C_{22,22}-2C_{12,21}+C_{11,22}+2C_{21,21}-2C_{32,23}-2C_{42,24}-C_{41,23}-2C_{31,24}\nonumber
\\
&-2C_{33,33}-2C_{43,34}+C_{44,33}+2C_{34,34}-2C_{13,31}-2C_{23,32}-2C_{24,31}-C_{14,32}\nonumber
\\
&-2C_{44,44}-2C_{34,43}+C_{33,44}+2C_{43,43}-2C_{24,42}-2C_{14,41}-2C_{13,42}-C_{23,41})~.
\label{eq_lebtotbub}
\end{align}

\section{Bethe-Salpeter equation for the static cooperon}
\label{subsec_Cooperon}

\subsection{General consideration}

We analyze the static Cooperon modes of $\widetilde{H}_{\textrm{LEB}}$ by considering the Bethe-Salpeter diagram in Figure~\ref{fig_manybody}(c). In the following sections, we assume that the scalar disorder potential dominates the elastic scattering, $|u_{00}|\gg|u_{ij}|$, and that the remaining symmetry-allowed channels of Eq.~(\ref{eq_DisOp}) are equally strong within each group, i.e. $\tau_{xz}=\tau_{y0}$ for the intravalley channels and $\tau_{0y}=\tau_{yy}=\tau_{zx}$ for the intervalley ones. This assumption is made for simplicity of calculation; since the logarithmic divergence is controlled by whether a Cooperon pole reaches $q=0$, relaxing the condition of equal disorder strength in each channel only changes the gap values, not the presence or absence of the singular correction. Using the Feynman rules, the Cooperon tensor $C$ satisfies
\begin{align}
\label{eq_abbrevCoop}
\left[C(\bm{q})\right]_{\alpha'\beta',\alpha\beta}=V_{\alpha'\alpha}V_{\beta'\beta}+\sum_{\gamma'\delta'}\left[C(\bm{q})\right]_{\alpha'\beta',\gamma'\delta'}\left[M(\bm{q})\right]_{\gamma'\delta',\alpha\beta}
\end{align}
with
\begin{align}
\label{eq_Melem}
\left[M(\bm{q})\right]_{\gamma'\delta',\alpha\beta}\equiv\sum_{\bm{k}}\sum_{\gamma\delta}\left[G_R(\bm{k})\right]_{\gamma'\gamma}\left[G_A(-\bm{k}+\bm{q})\right]_{\delta'\delta}
V_{\gamma\alpha}V_{\delta\beta}~.
\end{align}

Before computing the Cooperon modes, we investigate the product of Greens's function elements, $\left[G_R(\bm{q}_1)\right]_{\gamma'\gamma}\left[G_A(\bm{q}_2)\right]_{\delta'\delta}$. We start by rewriting the Eq.~(\ref{eq_GAV}) as
\begin{align}
G_{R/A}(\bm{p})=\frac{1}{D_{R/A}(\bm{p})}
\begin{pmatrix}
a_{R/A}(\bm{p})&b_{R/A}(\bm{p})&0&0\\
c_{R/A}(\bm{p})&a_{R/A}(\bm{p})&0&0\\
0&0&a_{R/A}(\bm{p})&-c_{R/A}(\bm{p})\\
0&0&-b_{R/A}(\bm{p})&a_{R/A}(\bm{p})
\end{pmatrix}~,
\end{align}
where
\begin{align}
&D_{R/A}(\bm{p})\equiv\epsilon_{R/A}^2-v^2p^2~,\nonumber
\\
&a_{R/A}(\bm{p})\equiv\epsilon_{R/A}~,\nonumber
\\
&b_{R/A}(\bm{p})\equiv vpe^{-i\theta_{\bm{p}}}~,\nonumber
\\
&c_{R/A}(\bm{p})\equiv vpe^{i\theta_{\bm{p}}}
\end{align}
with $pe^{i\theta_{\bm{p}}}\equiv p_x+ip_y$. Then, the possible products of the elements of retarded and advanced Green's function are only nine,
\begin{align}
&\frac{a_{R}(\bm{q}_1)a_{A}(\bm{q}_2)}{D_R(\bm{q}_1)D_A(\bm{q}_2)}~,\quad\pm\frac{a_{R}(\bm{q}_1)b_{A}(\bm{q}_2)}{D_R(\bm{q}_1)D_A(\bm{q}_2)}~,\quad\pm\frac{a_{R}(\bm{q}_1)c_{A}(\bm{q}_2)}{D_R(\bm{q}_1)D_A(\bm{q}_2)}~,\quad\pm\frac{b_{R}(\bm{q}_1)a_{A}(\bm{q}_2)}{D_R(\bm{q}_1)D_A(\bm{q}_2)}~,\quad\pm\frac{b_{R}(\bm{q}_1)b_{A}(\bm{q}_2)}{D_R(\bm{q}_1)D_A(\bm{q}_2)}~,\nonumber
\\
&\pm\frac{b_{R}(\bm{q}_1)c_{A}(\bm{q}_2)}{D_R(\bm{q}_1)D_A(\bm{q}_2)}~,\quad\pm\frac{c_{R}(\bm{q}_1)a_{A}(\bm{q}_2)}{D_R(\bm{q}_1)D_A(\bm{q}_2)}~,\quad\pm\frac{c_{R}(\bm{q}_1)b_{A}(\bm{q}_2)}{D_R(\bm{q}_1)D_A(\bm{q}_2)}~,\quad\pm\frac{c_{R}(\bm{q}_1)c_{A}(\bm{q}_2)}{D_R(\bm{q}_1)D_A(\bm{q}_2)}
\end{align}
and we only have to calculate these nine quantities, which are calculated are as below.
\begin{align}
\sum_{\bm{k}}\frac{a_{R}(\bm{k})a_{A}(-\bm{k}+\bm{q})}{D_R(\bm{k})D_A(-\bm{k}+\bm{q})}&=\sum_{\bm{k}}\frac{\epsilon_R}{\epsilon_R^2-v^2k^2}\frac{\epsilon_A}{\epsilon_A^2-v^2|-\bm{k}+\bm{q}|^2}\nonumber
\\
&\simeq\sum_{\bm{k}}\frac{\epsilon_R}{\epsilon_R^2-v^2k^2}\frac{\epsilon_A}{\epsilon_A^2-v^2k^2(1-q\cos{\phi}/k+q^2\sin^2{\phi}/2k^2)^2}\nonumber
\\
&\simeq\int\frac{d^2k}{4\pi^2}\frac{\epsilon_R\epsilon_A}{\epsilon_R^2-v^2k^2}\left[\frac{1}{\epsilon_A^2-v^2k^2}-\frac{2v^2kq\cos{\phi}}{(\epsilon_A^2-v^2k^2)^2}
+\frac{\epsilon_A^2v^2q^2+v^4k^2q^2+2v^4k^2q^2\cos2\phi}{(\epsilon_A^2-v^2k^2)^3}\right]\nonumber
\\
&=\int_0^{\infty}\frac{kdk}{2\pi}\underbrace{\frac{\epsilon_R\epsilon_A}{(v^2k^2-\epsilon_R^2)(v^2k^2-\epsilon_A^2)}}_\text{\encircle{A}}+\underbrace{\frac{\epsilon_R\epsilon_Av^2q^2(\epsilon_A^2+v^2k^2)}{(v^2k^2-\epsilon_R^2)(v^2k^2-\epsilon_A^2)^3}}_\text{\encircle{B}}~,
\end{align}
where $\phi$ is angle between $\bm{k}$ and $\bm{q}$. We first calculate $\encircle{A}$ ,
\begin{align}
\encircle{A}&=\int_0^{\infty}d\epsilon\frac{\epsilon}{2\pi v^2}\frac{\epsilon_R\epsilon_A}{(\epsilon^2-\epsilon_R^2)(\epsilon^2-\epsilon_A^2)}\qquad \left(x\equiv\epsilon^2\right)\nonumber
\\
&=\frac{\epsilon_R\epsilon_A}{4\pi v^2}\int_0^{\infty}dx\frac{1}{(x-\epsilon_R^2)(x-\epsilon_A^2)}\nonumber
\\
&\simeq\frac{E_F^2}{4\pi v^2}\int_{-\infty}^{\infty}dx\frac{1}{(x-\epsilon_R^2)(x-\epsilon_A^2)}\nonumber
\\
&=\frac{i}{4v^2}\frac{E_F^2}{2E_F\frac{i}{\tau_e}}\nonumber
\\
&=\frac{E_F\tau_e}{4v^2}~,
\end{align}
using $\epsilon_{R/A}\equiv E_F\pm i/2\tau_e$. Here, $\tau_e^{-1}$ is the scattering rate $\tau_{00}^{-1}$, $\tau_{00}^{-1}+\tau_{\textrm{intra}}^{-1}$, or $\tau_{00}^{-1}+\tau_{\textrm{inter}}^{-1}$, depending on the type of disorder potential, where
\begin{align}
&\tau_{\textrm{intra}}^{-1}\equiv\tau_{xz}^{-1}+\tau_{y0}^{-1}~,\nonumber
\\
&\tau_{\textrm{inter}}^{-1}\equiv\tau_{yy}^{-1}+\tau_{zx}^{-1}+\tau_{0y}^{-1}
\end{align}
with $\tau_{ij}^{-1}\equiv\pi N_0u_{ij}^2$. For $\encircle{B}$,
\begin{align}
\encircle{B}&=\int_0^{\infty}d\epsilon\frac{\epsilon}{2\pi v^2}\frac{\epsilon_R\epsilon_Av^2q^2(\epsilon_A^2+\epsilon^2)}{(\epsilon^2-\epsilon_R^2)(\epsilon^2-\epsilon_A^2)^3}~\qquad \left(x=\epsilon^2\right)\nonumber
\\
&=\frac{\epsilon_R\epsilon_Av^2q^2}{4\pi v^2}\int_0^{\infty}dx\frac{\epsilon_A^2+x}{(x-\epsilon_R^2)(x-\epsilon_A^2)^3}\nonumber
\\
&=\frac{\epsilon_R\epsilon_Av^2q^2}{4\pi v^2}2\pi i\frac{\epsilon_R^2+\epsilon_A^2}{(\epsilon_R^2-\epsilon_A^2)^3}\nonumber
\\
&\simeq\frac{iE_F^2v^2q^2}{2v^2}\frac{2E_F^2}{(2E_F)^3}\frac{1}{\left(\frac{i}{\tau_e}\right)^3}\nonumber
\\
&=-\frac{E_F\tau_e}{8v^2}v^2q^2\tau_e^2~,
\end{align}
where the last expression is obtained by keeping the leading term in $\tau_e$. Substituting $x=\epsilon^2=v^2k^2$, the measure becomes $\sum_k=\int_0^{\infty}dx/(4\pi v^2)$ and the integrand is rational in $x$ with poles only at $x=\epsilon_{R/A}^2$, both with $\Re x\simeq E_F^2>0$. The lower limit may therefore be extended to $-\infty$: the added region contains no pole and carries no $\tau$-enhancement, contributing a relative correction of order $(E_F\tau_e)^{-1}$ and $(E_F\tau_e)^{-3}$ for $\encircle{A}$ and $\encircle{B}$, respectively. Summing $\encircle{A}$ and $\encircle{B}$,
\begin{align}
\sum_{\bm{k}}\frac{a_{R}(\bm{k})a_{A}(-\bm{k}+\bm{q})}{D_R(\bm{k})D_A(-\bm{k}+\bm{q})}=\frac{E_F\tau_e}{2v^2}\left(\frac{1}{2}-\frac{1}{4}v^2q^2\tau_e^2\right)~.
\label{eq_aa}
\end{align}
We calculate the remaining 8 integrals likewise and summarize as below:
\begin{align}
&\sum_{\bm{k}}\frac{a_{R}(\bm{k})b_{A}(-\bm{k}+\bm{q})}{D_R(\bm{k})D_A(-\bm{k}+\bm{q})}=\frac{E_F\tau_e}{2v^2}\left(\frac{i}{4}e^{-i\eta}vq\tau_e\right)\equiv\frac{E_F\tau_e}{2v^2}iu^*~,\nonumber
\\
&\sum_{\bm{k}}\frac{a_{R}(\bm{k})c_{A}(-\bm{k}+\bm{q})}{D_R(\bm{k})D_A(-\bm{k}+\bm{q})}=\frac{E_F\tau_e}{2v^2}\left(\frac{i}{4}e^{i\eta}vq\tau_e\right)\equiv\frac{E_F\tau_e}{2v^2}iu~,\nonumber
\\
&\sum_{\bm{k}}\frac{b_{R}(\bm{k})a_{A}(-\bm{k}+\bm{q})}{D_R(\bm{k})D_A(-\bm{k}+\bm{q})}=\frac{E_F\tau_e}{2v^2}\left(-\frac{i}{4}e^{-i\eta}vq\tau_e\right)\equiv-\frac{E_F\tau_e}{2v^2}iu^*~,\nonumber
\\
&\sum_{\bm{k}}\frac{b_{R}(\bm{k})b_{A}(-\bm{k}+\bm{q})}{D_R(\bm{k})D_A(-\bm{k}+\bm{q})}=\frac{E_F\tau_e}{2v^2}\left(\frac{1}{8}e^{-2i\eta}v^2q^2\tau_e^2\right)\equiv\frac{E_F\tau_e}{2v^2}s^*~,\nonumber
\\
&\sum_{\bm{k}}\frac{b_{R}(\bm{k})c_{A}(-\bm{k}+\bm{q})}{D_R(\bm{k})D_A(-\bm{k}+\bm{q})}=\frac{E_F\tau_e}{2v^2}\left(-\frac{1}{2}+\frac{1}{4}v^2q^2\tau_e^2\right)\equiv-\frac{E_F\tau_e}{2v^2}p~,\nonumber
\\
&\sum_{\bm{k}}\frac{c_{R}(\bm{k})a_{A}(-\bm{k}+\bm{q})}{D_R(\bm{k})D_A(-\bm{k}+\bm{q})}=\frac{E_F\tau_e}{2v^2}\left(-\frac{i}{4}e^{i\eta}vq\tau\right)\equiv-\frac{E_F\tau_e}{2v^2}iu~,\nonumber
\\
&\sum_{\bm{k}}\frac{c_{R}(\bm{k})b_{A}(-\bm{k}+\bm{q})}{D_R(\bm{k})D_A(-\bm{k}+\bm{q})}=\frac{E_F\tau_e}{2v^2}\left(-\frac{1}{2}+\frac{1}{4}v^2q^2\tau_e^2\right)\equiv-\frac{E_F\tau_e}{2v^2}p~,\nonumber
\\
&\sum_{\bm{k}}\frac{c_{R}(\bm{k})c_{A}(-\bm{k}+\bm{q})}{D_R(\bm{k})D_A(-\bm{k}+\bm{q})}=\frac{E_F\tau_e}{2v^2}\left(\frac{1}{8}e^{2i\eta}v^2q^2\tau_e^2\right)\equiv\frac{E_F\tau_e}{2v^2}s~,
\label{eq_Mingredients}
\end{align}
where
\begin{align}
&p\equiv\frac{1}{2}-\frac{1}{4}v^2q^2\tau_e^2~,\nonumber
\\
&u\equiv\frac{1}{4}e^{i\eta}vq\tau_e~,\nonumber
\\
&s\equiv\frac{1}{8}e^{2i\eta}v^2q^2\tau_e^2
\end{align}
and $\tan\eta\equiv q_y/q_x$. These equations are ingredients for calculating the elements of $M$ for scalar, intravalley and intervalley disorder potentials.

\subsection{Scalar disorder potential case}

To proceed, we first calculate the Cooperon mode $C_{\alpha'\beta',\alpha\beta}$ when only the scalar disorder potential exists. As a representative example, we consider $C_{11,11}$, $C_{11,12}$, $C_{11,21}$ and $C_{11,22}$. From Eq.~(\ref{eq_abbrevCoop}), we see that $4^2$ Cooperon modes are needed to calculate $C_{11,11}$,
\begin{align}
C_{11,11}=u_{00}^2&+C_{11,11}M_{11,11}+C_{11,12}M_{12,11}+C_{11,13}M_{13,11}+C_{11,14}M_{14,11}+C_{11,21}M_{21,11}+C_{11,22}M_{22,11}\nonumber
\\
&+C_{11,23}M_{23,11}+C_{11,24}M_{24,11}+C_{11,31}M_{31,11}+C_{11,32}M_{32,11}+C_{11,33}M_{33,11}+C_{11,34}M_{34,11}\nonumber
\\
&+C_{11,41}M_{41,11}+C_{11,42}M_{42,11}+C_{11,43}M_{43,11}+C_{11,44}M_{44,11}~.
\label{eq_c11eq}
\end{align}
However, since the scalar disorder potential that induces the scattering within a single valley, the matrix elements of $M$ that include the indices $3$ and $4$ are zero. Therefore, Eq.~(\ref{eq_c11eq}) reduces to
\begin{align}
C_{11,11}=u_{00}^2+C_{11,11}M_{11,11}+C_{11,12}M_{12,11}+C_{11,21}M_{21,11}+C_{11,22}M_{22,11}~,
\end{align}
where
\begin{align}
M_{\gamma'\delta',11}=u_{00}^2\left[G_R(\bm{k})\right]_{\gamma'1}\left[G_A(-\bm{k}+\bm{q})\right]_{\delta'1}~.
\label{eq_Mscalar}
\end{align}
When we further apply this logic to $C_{11,12}$, $C_{11,21}$ and $C_{11,2}$, we obtain the matrix equation
\begin{align}
\begin{pmatrix}
C_{11,11}\\C_{11,12}\\C_{11,21}\\C_{11,22}
\end{pmatrix}
=u_{00}^2
\begin{pmatrix}
1\\0\\0\\0
\end{pmatrix}
+
\begin{pmatrix}
M_{11,11}&M_{12,11}&M_{21,11}&M_{22,11}
\\
M_{11,12}&M_{12,12}&M_{21,12}&M_{22,12}
\\
M_{11,21}&M_{12,21}&M_{21,21}&M_{22,21}
\\
M_{11,22}&M_{12,22}&M_{21,22}&M_{22,22}
\end{pmatrix}
\begin{pmatrix}
C_{11,11}\\C_{11,12}\\C_{11,21}\\C_{11,22}
\end{pmatrix}~.
\label{eq_Cmatrixeq}
\end{align}
Since we are considering the scalar disorder potential only, $\tau_e=\tau_{00}$ and $\tau_{00}^{-1}=\pi N_0u_{00}^2$. Thus, the factor $E_F\tau_e/2v^2$ in Eq.~(\ref{eq_Mingredients}) combined with the square of the scalar disorder potential strength $u_{00}^2$ yields
\begin{align}
u_{00}^2\frac{E_F\tau_{00}}{2v^2}=1.
\label{eq_scalarWard}
\end{align}
Thus using Eq.~(\ref{eq_Mscalar}) and Eqs.~(\ref{eq_aa})--(\ref{eq_Mingredients}), Eq.~(\ref{eq_Cmatrixeq}) is written as
\begin{align}
\begin{pmatrix}
C_{11,11}\\C_{11,12}\\C_{11,21}\\C_{11,22}
\end{pmatrix}
=u_{00}^2
\begin{pmatrix}
1\\0\\0\\0
\end{pmatrix}
+
\begin{pmatrix}
p&iu&-iu&s
\\
iu^*&p&-p&-iu
\\
-iu^*&-p&p&iu
\\
s^*&-iu^*&iu^*&p
\end{pmatrix}
\begin{pmatrix}
C_{11,11}\\C_{11,12}\\C_{11,21}\\C_{11,22}
\end{pmatrix}
\label{eq_c11}
\end{align}
with
\begin{align}
p=\frac{1}{2}-\frac{1}{4}v^2q^2\tau_{00}^2~,\quad u=\frac{1}{4}e^{i\eta}vq\tau_{00}~,\quad s=\frac{1}{8}e^{2i\eta}v^2q^2\tau_{00}^2~.
\end{align}
We solve Eq.~(\ref{eq_c11}) and obtain
\begin{align}
&C_{11,11}=\frac{16(3+v^2q^2\tau_{00}^2)u_{00}^2}{(4+v^2q^2\tau_{00}^2)(8+3v^2q^2\tau_{00}^2)}\simeq\frac{3}{2}u_{00}^2~,\nonumber
\\
&C_{11,12}=\frac{4ie^{-i\eta}u_{00}^2}{vq\tau_{00}(8+3v^2q^2\tau_{00}^2)}\simeq\frac{ie^{-i\eta}}{2vq\tau_{00}}u_{00}^2~,\nonumber
\\
&C_{11,21}=-\frac{4ie^{-i\eta}u_{00}^2}{vq\tau_{00}(8+3v^2q^2\tau_{00}^2)}\simeq-\frac{ie^{-i\eta}}{2vq\tau_{00}}u_{00}^2~,\nonumber
\\
&C_{11,22}=\frac{8e^{-i2\eta}(2+v^2q^2\tau_{00}^2)u_{00}^2}{(4+v^2q^2\tau_{00}^2)(8+3v^2q^2\tau_{00}^2)}\simeq\frac{e^{-2i\eta}}{2}u_{00}^2~.
\end{align}
Similarly, we write the matrix equation that $C_{12,11}$, $C_{12,12}$, $C_{12,21}$ and $C_{12,22}$ satisfy,
\begin{align}
\begin{pmatrix}
C_{12,11}\\C_{12,12}\\C_{12,21}\\C_{12,22}
\end{pmatrix}
=u_{00}^2
\begin{pmatrix}
0\\1\\0\\0
\end{pmatrix}
+
\begin{pmatrix}
p&iu&-iu&s
\\
iu^*&p&-p&-iu
\\
-iu^*&-p&p&iu
\\
s^*&-iu^*&iu^*&p
\end{pmatrix}
\begin{pmatrix}
C_{12,11}\\C_{12,12}\\C_{12,21}\\C_{12,22}
\end{pmatrix}~.
\label{eq_matrixeqc12}
\end{align}
This equation gives
\begin{align}
&C_{12,11}\simeq\frac{ie^{i\theta}}{2vq\tau_{00}}u_{00}^2~,\nonumber
\\
&C_{12,12}\simeq\frac{1}{2v^2q^2\tau_{00}^2}u_{00}^2~,\nonumber
\\
&C_{12,21}\simeq-\frac{1}{2v^2q^2\tau_{00}^2}u_{00}^2~,\nonumber
\\
&C_{12,22}\simeq-\frac{ie^{-i\theta}}{2vq\tau_{00}}u_{00}^2~.
\label{eq_c12}
\end{align}
Among the Cooperon modes in Eq.~(\ref{eq_c11}) and Eq.~(\ref{eq_c12}), only $C_{11,11}$, $C_{11,22}$, $C_{12,12}$ and $C_{12,21}$ appear in Eq.~(\ref{eq_lebtotbub}), and these four Cooperon modes can be grouped into two, finite and diverging Cooperon modes. $C_{11,11}$ and $C_{11,22}$ are finite Cooperon modes, meaning that they give finite values in $q\rightarrow0$ limit. On the other hand, $C_{12,12}$ and $C_{12,21}$ are diverging Cooperon modes that have pole at $q=0$. By repeating the procedure we used to get Eq.~(\ref{eq_c11}) and Eq.~(\ref{eq_c12}), we group the Cooperon modes appearing in Eq.~(\ref{eq_lebtotbub}) into three, namely, finite, diverging and vanishing Cooperon modes,
\begin{alignat}{2}
&\textrm{Finite Cooperon modes: }
&&C_{11,11},C_{22,22},C_{33,33},C_{44,44},C_{11,22},C_{33,44},C_{22,11},C_{44,33}~,\nonumber
\\
&\textrm{Diverging Cooperon modes: }
&&C_{12,12},C_{21,21},C_{34,34},C_{43,43},C_{12,21},C_{21,12},C_{34,43},C_{43,34}~,\nonumber
\\
&\textrm{Vanishing Cooperon modes: }
&&C_{14,41},C_{14,32},C_{13,42},C_{13,31},C_{24,42},C_{24,31},C_{23,41},C_{23,32},\nonumber
\\
& &&C_{31,13},C_{31,24},C_{32,14},C_{32,23},C_{41,14},C_{41,23},C_{42,13},C_{42,24}~.
\label{eq_groupCoop}
\end{alignat}
In the static long-wavelength limit $q\rightarrow0$, the most singular divergence comes from the diverging Cooperon modes. Therefore, we can ignore finite Cooperon modes in the following arguments. Also, we stress that vanishing Cooperon modes above can take non-zero values when intervalley disorder potentials are present. Keeping diverging Cooperons only, Eq.~(\ref{eq_lebtotbub}) reduces to
\begin{align}
\delta g_x^{\textrm{scalar}}\simeq e^2N_0v^2\tau_{00}^3\sum_{\bm{q}}(-2C_{21,12}+2C_{12,12}-2C_{12,21}+2C_{21,21}-2C_{43,34}+2C_{34,34}-2C_{34,43}+2C_{43,43})~.
\label{eq_lebtotbubfin}
\end{align}

\subsection{Intravalley disorder potential case}

We next do the similar calculations to obtain the Cooperon modes appearing in Eq.~(\ref{eq_lebtotbubfin}) when intravalley disorder potentials, $u_{xz}\Sigma_x\Lambda_z$ and $u_{y0}\Sigma_y\Lambda_0$, are additionally applied. In this case, the scattering rate changes from $\tau_{00}^{-1}$ to $\tau_{\textrm{intra}}^{-1}$, and the matrix equation Eq.~(\ref{eq_Cmatrixeq}) is still valid without intervalley disorder potentials. However, the specific form of the elements of $M$ changes because now the system has 3 disorder potentials, $u_{00}\Sigma_0\Lambda_0$, $u_{xz}\Sigma_x\Lambda_z$ and $u_{y0}\Sigma_y\Lambda_0$. For instance, $M_{11,11}$ is calculated as
\begin{align}
M_{11,11}&=\sum_{\bm{k},\gamma,\delta}\left[G_R(\bm{k})\right]_{1\gamma}\left[G_A(-\bm{k}+\bm{q})\right]_{1\delta}\left[u_{00}^2(\Sigma_0\Lambda_0)_{\gamma1}(\Sigma_0\Lambda_0)_{\delta1}+u_{xz}^2(\Sigma_x\Lambda_z)_{\gamma1}(\Sigma_x\Lambda_z)_{\delta1}+u_{y0}^2(\Sigma_y\Lambda_0)_{\gamma1}(\Sigma_y\Lambda_0)_{\delta1}\right]\nonumber
\\
&=\sum_{\bm{k}}u_{00}^2\left[G_R(\bm{k})\right]_{11}\left[G_A(-\bm{k}+\bm{q})\right]_{11}+\left(u_{xz}^2-u_{y0}^2\right)\left[G_R(\bm{k})\right]_{12}\left[G_A(-\bm{k}+\bm{q})\right]_{12}\nonumber
\\
&=\sum_{\bm{k}}u_{00}^2\left[G_R(\bm{k})\right]_{11}\left[G_A(-\bm{k}+\bm{q})\right]_{11}
\end{align}
because $(\Sigma_x\Lambda_z)_{12}^2=(\Sigma_x\Lambda_z)_{21}^2=+1$ and $(\Sigma_y\Lambda_0)_{12}^2=(\Sigma_y\Lambda_0)_{21}^2=-1$ and we are assuming $u_{xz}=u_{y0}$. Using Eq.~(\ref{eq_aa}), $M_{11,11}$ becomes
\begin{align}
M_{11,11}=u_{00}^2\frac{E_F\tau}{2v^2}\left(\frac{1}{2}-\frac{1}{4}v^2q^2\tau^2\right)~,
\end{align}
where $\tau_{\textrm{intra}}^{-1}\equiv\tau_{xz}^{-1}+\tau_{y0}^{-1}$ and $\tau^{-1}\equiv\tau_{00}^{-1}+\tau_{\textrm{intra}}^{-1}$. When additional disorder potentials other than the scalar disorder potential exist, $u_{00}^2E_F\tau/2v^2\neq1$. Instead, have
\begin{align}
u_{00}^2\frac{E_F\tau}{2v^2}&=\frac{1}{\pi N_0\tau_{00}}\frac{2\pi v^2N_0\tau}{2v^2}\nonumber
\\
&=\frac{\tau}{\tau_{00}}\nonumber
\\
&=\tau\left(\tau^{-1}-\tau_{\textrm{intra}}^{-1}\right)\nonumber
\\
&\equiv1-r_{\textrm{intra}}~,
\end{align}
where $r_{\textrm{intra}}\equiv\tau/\tau_{\textrm{intra}}\ll1$. Thus, we obtain
\begin{align}
M_{11,11}=(1-r_{\textrm{intra}})\left(\frac{1}{2}-\frac{1}{4}v^2q^2\tau^2\right)~.
\end{align}
The cancellation of $u_{xz}^2$ and $u_{y0}^2$ occurs for $M_{12,11}$, $M_{21,11}$, $M_{22,11}$, $M_{11,22}$, $M_{12,22}$, $M_{21,22}$ and $M_{22,22}$ as well, so that they obtain the $(1-r_{\textrm{intra}})$ factor compared to the value of the scalar disorder potential case. On the other hand, the remaining 8 elements, $M_{\gamma'\delta',12}$ and $M_{\gamma'\delta',21}$, should be treated differently. As a representative example, $M_{11,12}$ becomes
\begin{align}
M_{11,12}&=\sum_{\bm{k}}
\sum_{\gamma,\delta}\left[G_R(\bm{k})\right]_{1\gamma}\left[G_A(-\bm{k}+\bm{q})\right]_{1\delta}\left[u_{00}^2(\Sigma_0\Lambda_0)_{\gamma1}(\Sigma_0\Lambda_0)_{\delta2}+u_{xz}^2(\Sigma_x\Lambda_z)_{\gamma1}(\Sigma_x\Lambda_z)_{\delta2}+u_{y0}^2(\Sigma_y\Lambda_0)_{\gamma1}(\Sigma_y\Lambda_0)_{\delta2}\right]\nonumber
\\
&=\sum_{\bm{k}}u_{00}^2\left[G_R(\bm{k})\right]_{11}\left[G_A(-\bm{k}+\bm{q})\right]_{12}+\left(u_{xz}^2+u_{y0}^2\right)\left[G_R(\bm{k})\right]_{12}\left[G_A(-\bm{k}+\bm{q})\right]_{11}\nonumber
\\
&=\left(u_{00}^2-u_{xz}^2-u_{y0}^2\right)\frac{E_F\tau}{2v^2}\left(\frac{i}{4}e^{-i\eta}vq\tau\right)\nonumber
\\
&=\tau\left(\tau_{00}^{-1}-\tau_{xz}^{-1}-\tau_{y0}^{-1}\right)\left(\frac{i}{4}e^{-i\eta}vq\tau\right)\nonumber
\\
&=\tau\left(\tau^{-1}-2\tau_{xz}^{-1}-2\tau_{y0}^{-1}\right)\left(\frac{i}{4}e^{-i\eta}vq\tau\right)\nonumber
\\
&=(1-2r_{\textrm{intra}})\left(\frac{i}{4}e^{-i\eta}vq\tau\right)~.
\end{align}
Similarly, $M_{12,12}$, $M_{21,12}$. $M_{22,12}$, $M_{11,21}$, $M_{12,21}$, $M_{21,21}$, $M_{22,21}$, $M_{11,22}$, $M_{12,22}$, $M_{21,22}$ and $M_{22,22}$ obtain the $(1-2r_{\textrm{intra}})$ factor compared to the value of the scalar disorder potential case. Thus, the matrix equation that $C_{12,11}$, $C_{12,12}$, $C_{12,21}$ and $C_{12,22}$ satisfy changes from Eq.~(\ref{eq_matrixeqc12}) to
\begin{align}
\begin{pmatrix}
C_{12,11}\\C_{12,12}\\C_{12,21}\\C_{12,22}
\end{pmatrix}
=u_{00}^2
\begin{pmatrix}
0\\1\\0\\0
\end{pmatrix}
+
\begin{pmatrix}
(1-r_{\textrm{intra}})p&(1-r_{\textrm{intra}})iu&-(1-r_{\textrm{intra}})iu&(1-r_{\textrm{intra}})s
\\
(1-2r_{\textrm{intra}})iu^*&(1-2r_{\textrm{intra}})p&-(1-2r_{\textrm{intra}})p&-(1-2r_{\textrm{intra}})iu
\\
-(1-2r_{\textrm{intra}})iu^*&-(1-2r_{\textrm{intra}})p&(1-2r_{\textrm{intra}})p&(1-2r_{\textrm{intra}})iu
\\
(1-r_{\textrm{intra}})s^*&-(1-r_{\textrm{intra}})iu^*&(1-r_{\textrm{intra}})iu^*&(1-r_{\textrm{intra}})p
\end{pmatrix}
\begin{pmatrix}
C_{12,11}\\C_{12,12}\\C_{12,21}\\C_{12,22}
\end{pmatrix}~,
\end{align}
where
\begin{align}
p=\frac{1}{2}-\frac{1}{4}v^2q^2\tau^2~,\quad u=\frac{i}{4}e^{i\eta}vq\tau~,\quad s=\frac{1}{8}e^{-2i\eta}v^2q^2\tau^2~.
\end{align}
The Cooperon modes that diverge in the scalar disorder potential case, $C_{12,12}$ and $C_{12,21}$, now become finite by obtaining the gap proportional to $r_{\textrm{intra}}$,
\begin{align}
&C_{12,12}\simeq\frac{1}{2v^2q^2\tau^2+4r_{\textrm{intra}}}u_{00}^2~,\nonumber
\\
&C_{12,21}\simeq-\frac{1}{2v^2q^2\tau^2+4r_{\textrm{intra}}}u_{00}^2~.
\label{eq_c12intra}
\end{align}
While there is no diverging Cooperon modes when the intravalley disorder potentials are applied, the leading contribution to the conductivity correction of the intravalley disorder potential case has the same form as Eq.~(\ref{eq_lebtotbubfin}),
\begin{align}
\delta g_x^{\textrm{intra}}\simeq e^2N_0v^2\tau_{\textrm{intra}}^3\sum_{\bm{q}}(-2C_{21,12}+2C_{12,12}-2C_{12,21}+2C_{21,21}-2C_{43,34}+2C_{34,34}-2C_{34,43}+2C_{43,43})~.
\label{eq_intralebtotbub}
\end{align}

\subsection{Intervalley disorder potential case}

We next apply the scalar disorder potential and intervalley disorder potentials $u_{yy}\Sigma_y\Lambda_y$, $u_{zx}\Sigma_z\Lambda_x$ and $u_{0y}\Sigma_0\Lambda_y$. Since the scattering between two valleys are activated due to intervalley disorder potentials, the matrix elements of $M$ that involve different valleys do not vanish. According to Eq.~(\ref{eq_abbrevCoop}), $C_{12,11}$ is determined from
\begin{align}
C_{12,11}=&C_{12,11}M_{11,11}+C_{12,12}M_{12,11}+C_{12,21}M_{21,11}+C_{12,22}M_{22,11}\nonumber
\\
&+C_{12,33}M_{33,11}+C_{12,34}M_{34,11}+C_{12,43}M_{43,11}+C_{12,44}M_{44,11}~.
\label{eq_c12inter}
\end{align}
Therefore, we should solve the $8\times8$ matrix equation, not $4\times4$. As in the intravalley disorder potential case, additional factors emerge. Specifically, as a representative example, we consider
\begin{align}
M_{11,11} &= \sum_{\bm{k},\gamma,\delta}\left[G_R(\bm{k})\right]_{1\gamma}\left[G_A(-\bm{k}+\bm{q})\right]_{1\delta}\left[u_{00}^2(\Sigma_0\Lambda_0)_{\gamma1}(\Sigma_0\Lambda_0)_{\delta1}+u_{yy}^2(\Sigma_y\Lambda_y)_{\gamma1}(\Sigma_y\Lambda_y)_{\delta1}\right.~\nonumber
\\
&\phantom{= \sum_{\bm{k},\gamma,\delta}\left[G_R(\bm{k})\right]_{1\gamma}\left[G_A(-\bm{k}+\bm{q})\right]_{1\delta}} \left.+u_{zx}^2(\Sigma_z\Lambda_x)_{\gamma1}(\Sigma_z\Lambda_x)_{\delta1}+u_{0y}^2(\Sigma_0\Lambda_y)_{\gamma1}(\Sigma_0\Lambda_y)_{\delta1}\right]\nonumber
\\
&= \sum_{\bm{k}}u_{00}^2\left[G_R(\bm{k})\right]_{11}\left[G_A(-\bm{k}+\bm{q})\right]_{11}+u_{yy}^2\left[G_R(\bm{k})\right]_{14}\left[G_A(-\bm{k}+\bm{q})\right]_{14}+\left(u_{zx}^2-u_{0y}^2\right)\left[G_R(\bm{k})\right]_{13}\left[G_A(-\bm{k}+\bm{q})\right]_{13}\nonumber
\\
&= \sum_{\bm{k}}u_{00}^2\left[G_R(\bm{k})\right]_{11}\left[G_A(-\bm{k}+\bm{q})\right]~,\nonumber
\\
&=u_{00}^2\frac{E_F\tau}{2v^2}\left(\frac{1}{2}-\frac{1}{4}v^2q^2\tau^2\right)\nonumber
\\
&=\frac{\tau}{\tau_{00}}\left(\frac{1}{2}-\frac{1}{4}v^2q^2\tau^2\right)\nonumber
\\
&=\tau\left(\tau^{-1}-\tau_{yy}^{-1}-\tau_{zx}^{-1}-\tau_{0y}^{-1}\right)\left(\frac{1}{2}-\frac{1}{4}v^2q^2\tau^2\right)\nonumber
\\
&\equiv(1-r_{\textrm{inter}})\left(\frac{1}{2}-\frac{1}{4}v^2q^2\tau^2\right)~,
\end{align}
where $\tau^{-1}\equiv\tau_{00}^{-1}+\tau_{\textrm{inter}}^{-1}$, $\tau_{\textrm{inter}}\equiv\tau_{yy}^{-1}+\tau_{zx}^{-1}+\tau_{0y}^{-1}$, and $r_{\textrm{inter}}\equiv\tau/\tau_{\textrm{inter}}$. The $(1-r_{\textrm{inter}})$ factor emerges because Green's function is valley-diagonal, $(G_{R/A})_{13}=(G_{R/A})_{14}=0$, and $M_{12,11}$, $M_{21,11}$ and $M_{22,11}$ in Eq.~(\ref{eq_c12inter}) acquire the $(1-r_{\textrm{inter}})$ factor as well. On the other hand, the elements of $M$ involving intervalley scattering are different. For instance,
\begin{align}
M_{33,11} &= \sum_{\bm{k},\gamma,\delta}\left[G_R(\bm{k})\right]_{3\gamma}\left[G_A(-\bm{k}+\bm{q})\right]_{3\delta}\left[u_{00}^2(\Sigma_0\Lambda_0)_{\gamma1}(\Sigma_0\Lambda_0)_{\delta1}+u_{yy}^2(\Sigma_y\Lambda_y)_{\gamma1}(\Sigma_y\Lambda_y)_{\delta1}\right.~\nonumber
\\
&\phantom{= \sum_{\bm{k},\gamma,\delta}\left[G_R(\bm{k})\right]_{3\gamma}\left[G_A(-\bm{k}+\bm{q})\right]_{3\delta}} \left.+u_{zx}^2(\Sigma_z\Lambda_x)_{\gamma1}(\Sigma_z\Lambda_x)_{\delta1}+u_{0y}^2(\Sigma_0\Lambda_y)_{\gamma1}(\Sigma_0\Lambda_y)_{\delta1}\right]\nonumber
\\
&=\sum_{\bm{k}}u_{yy}^2\left[G_R(\bm{k})\right]_{34}\left[G_A(-\bm{k}+\bm{q})\right]_{34}+\left(u_{zx}^2-u_{0y}^2\right)\left[G_R(\bm{k})\right]_{33}\left[G_A(-\bm{k}+\bm{q})\right]_{33}\nonumber
\\
&=\sum_{\bm{k}}u_{yy}^2\left[G_R(\bm{k})\right]_{34}\left[G_A(-\bm{k}+\bm{q})\right]_{34}\nonumber
\\
&=u_{yy}^2\frac{E_F\tau}{2v^2}\left(\frac{1}{8}e^{2i\eta}v^2q^2\tau^2\right)\nonumber
\\
&=\frac{\tau}{\tau_{yy}}\left(\frac{1}{8}e^{2i\eta}v^2q^2\tau^2\right)\nonumber
\\
&=\frac{r_{\textrm{inter}}}{3}\left(\frac{1}{8}e^{2i\eta}v^2q^2\tau^2\right)~.
\end{align}
After calculating the other elements following the similar procedure, $8\times8$ matrix equation is written as
\begin{align}
\begin{pmatrix}
C_{12,11}
\\
C_{12,12}
\\
C_{12,21}
\\
C_{12,22}
\\
C_{12,33}
\\
C_{12,34}
\\
C_{12,43}
\\
C_{12,44}
\end{pmatrix}
\simeq
u_{00}^2
\begin{pmatrix}
0
\\
1
\\
0
\\
0
\\
0
\\
0
\\
0
\\
0
\end{pmatrix}
+
\begin{pmatrix}
T_{11}&T_{12}
\\
T_{21}&T_{22}
\end{pmatrix}
\begin{pmatrix}
C_{12,11}
\\
C_{12,12}
\\
C_{12,21}
\\
C_{12,22}
\\
C_{12,33}
\\
C_{12,34}
\\
C_{12,43}
\\
C_{12,44}
\end{pmatrix}~,
\label{eq_matc12inter}
\end{align}
where
\begin{align}
&T_{11}\equiv(1-r_{\textrm{inter}})
\begin{pmatrix}
p&iu&-iu&s
\\
iu^*&p&-p&-iu
\\
-iu^*&-p&p&iu
\\
s^*&-iu^*&iu^*&p
\end{pmatrix}~,\nonumber
\\
&T_{12}\equiv\frac{r_{\textrm{inter}}}{3}
\begin{pmatrix}
s&iu&-iu&p
\\
-iu&p&-p&iu^*
\\
iu&-p&p&-iu^*
\\
p&-iu^*&iu^*&s^*
\end{pmatrix}~,\nonumber
\\
&T_{21}\equiv\frac{r_{\textrm{inter}}}{3}
\begin{pmatrix}
s^*&-iu^*&iu^*&p
\\
iu^*&p&-p&-iu
\\
-iu^*&-p&p&iu
\\
p&iu&-iu&s
\end{pmatrix}~,\nonumber
\\
&T_{22}\equiv(1-r_{\textrm{inter}})
\begin{pmatrix}
p&-iu^*&iu^*&s^*
\\
-iu&p&-p&iu^*
\\
iu&-p&p&-iu^*
\\
s&iu&-iu&p
\end{pmatrix}~.
\label{eq_interTmat}
\end{align}
We here assume that $u_{00}\gg u_{ij}$ so that $u_{00}$ dominates the first term on the right-hand side in Eq.~(\ref{eq_abbrevCoop}). We obtain $C_{12,12}$ and $C_{12,21}$ that used to diverge for the scalar disorder potential case by solving Eq.~(\ref{eq_matc12inter}),
\begin{align}
&C_{12,12}\simeq\frac{1}{2v^2q^2\tau^2+4r_{\textrm{inter}}}u_{00}^2~,\nonumber
\\
&C_{12,21}\simeq-\frac{1}{2v^2q^2\tau^2+4r_{\textrm{inter}}}u_{00}^2~.
\end{align}
Again, these Cooperon modes acquire a gap proportional to $r_{\textrm{inter}}$ and no longer diverge.

Singular Cooperon modes in the intervalley disorder potential case originate from the vanishing Cooperon modes in the scalar and the intravalley disorder potential cases, Eq.~(\ref{eq_groupCoop}). We calculate these Cooperon modes following the procedure we used in Eqs.~(\ref{eq_c12inter})--(\ref{eq_interTmat}) and obtain
\begin{align}
\{C_{13,42},C_{13,31},C_{24,42},C_{24,31},C_{31,13},C_{31,24},C_{42,13},C_{42,24}\}\simeq\frac{1}{4v^2q^2\tau^2}u_{00}^2
\label{eq_gaplessinter}
\end{align}
for $0<vq\tau\ll r_{\textrm{inter}}$. All Cooperon modes other than Eq.~(\ref{eq_gaplessinter}) are finite. Therefore, keeping only the diverging Cooperon modes, Eq.~(\ref{eq_lebtotbub}) for the intervalley disorder potential case becomes
\begin{align}
\delta g_x^{\textrm{inter}}\simeq e^2N_0v^2\tau^3\sum_{\bm{q}}(&-2C_{31,13}-2C_{42,13}-2C_{42,24}-2C_{31,24}-2C_{13,31}-2C_{24,31}-2C_{24,42}-2C_{13,42})~.
\label{eq_interlebtotbub}
\end{align}

\section{The Quantum Correction To The Conductivity of the Lower Euler bands}
\label{supsec_dg}

\subsection{The LEB with inversion symmetry $(\lambda_R=0)$}

In this section, we evaluate the DC quantum correction to the conductivity due to disorder using the Cooperons obtained in Sec.~\ref{subsec_Cooperon}. We begin with the LEB under $\lambda_R=0$ and $\lambda_m\neq0$ in the presence of the scalar disorder potential only. Plugging the analytic expressions of diverging Cooperon modes of Eq.~(\ref{eq_groupCoop}) into Eq.~(\ref{eq_lebtotbubfin}), we obtain
\begin{align}
\delta g_x^{\textrm{scalar}}&\simeq e^2N_0v^2\tau_{00}^3\sum_{\bm{q}}\left(16\frac{u_{00}^2}{2v^2q^2\tau_{00}^2}\right)\nonumber
\\
&=\frac{8e^2}{\pi}\int\frac{d^2q}{4\pi^2}\frac{1}{q^2}\nonumber
\\
&=\frac{4e^2}{\pi^2}\int_{(v^2\tau_{00}\tau_{\phi})^{-\frac{1}{2}}}^{(v^2\tau_{00}^2)^{-\frac{1}{2}}}dq\frac{1}{q}\nonumber
\\
&=\frac{2e^2}{\pi^2}\ln\frac{\tau_{\phi}}{\tau_{00}}~,
\end{align}
where $\tau_{\phi}^{-1}$ is the inelastic scattering rate of electrons. The limits of the $q$ integral are set by the two length scales that bound the diffusive regime: the lower limit $q_{\textrm{min}}=(v^2\tau_{00}\tau_{\phi})^{-1/2}$ is the inverse of the dephasing length $L_{\phi}\equiv\sqrt{F\tau_{\phi}}$ with the diffusion constant $F\equiv v^2\tau_{00}$, beyond which phase coherence is lost, while the upper limit $q_{\textrm{max}}=(v^2\tau_{00}^2)^{-1/2}$ is the inverse mean free path, below which the diffusive description ceases to apply. We find that the leading term has the logarithmic divergence, $\ln\tau_{\phi}/\tau_{00}$, in the limit $\tau_{\phi}\rightarrow\infty$. Since $\tau_{00}^{-1}\gg \tau_{\phi}^{-1}$, the logarithm dominates over the negative constant terms coming from the finite Cooperons and $\delta g_x^{\textrm{scalar}}$ is positive, i.e. of the WAL type.

For the intravalley disorder potential case, we calculate the leading contribution of conductivity correction using Eq.~(\ref{eq_c12intra}) and Eq.~(\ref{eq_intralebtotbub}),
\begin{align}
\delta g_x^{\textrm{intra}}&\simeq e^2N_0v^2\tau^3\sum_{\bm{q}}\left(16\frac{u_{00}^2}{2v^2q^2\tau^2+4r_{\textrm{intra}}}\right)\nonumber
\\
&\simeq\frac{8e^2}{\pi}\int\frac{d^2q}{4\pi^2}\frac{v^2\tau^2}{v^2q^2\tau^2+2r_{\textrm{intra}}}\nonumber
\\
&=\frac{4e^2}{\pi^2}\int_{(v^2\tau\tau_{\phi})^{-\frac{1}{2}}}^{(v^2\tau^2)^{-\frac{1}{2}}}dq\frac{v^2q\tau^2}{v^2q^2\tau^2+2r_{\textrm{intra}}}\nonumber
\\
&=\frac{2e^2}{\pi^2}\ln\frac{2r_{\textrm{intra}}+1}{2r_{\textrm{intra}}+\tau/\tau_{\phi}}~.
\label{eq_findgx2}
\end{align}
Since Eq.~(\ref{eq_findgx2}) remains finite as $\tau_{\phi}\rightarrow\infty$, the intravalley disorder potentials suppress the logarithmic quantum correction, in contrast to the scalar disorder potential case where the $r_{\textrm{intra}}\rightarrow0$ limit recovers the log divergence.

\begin{table}
\caption{{\bf Representative Cooperon modes that give singular contribution in leading order and how each disorder potential affects the Cooperon modes.} The scalar disorder potential induces the diverging Cooperons, while making Cooperon modes that require the intervalley scattering vanish. When intravalley disorder potentials are additionally applied, Cooperon modes that previously diverge now obtain the finite gap proportional to $r_{\textrm{intra}}$ and become finite. Cooperon modes that vanish in scalar disorder potential case remain the same. On the other hand, when the intervalley disorder potentials are applied, gapped Cooperon modes are still gapped with the gap proportional to $r_{\textrm{inter}}$. However, Cooperon modes that previously vanish now become nonzero and show the singular behavior at $q=0$, which makes the leading conductivity correction diverges.}  
\begin{tabular}{c|ccc}
\hline\hline
 &\multicolumn{3}{c}{Disorder}\\ Cooperon&Scalar&Scalar+Intravalley&Scalar+Intervalley\\ \hline
$C_{12,12}$&$\frac{u_{00}^2}{2q^2v^2\tau_{00}^2}$&$\frac{u_{00}^2}{2q^2v^2\tau^2+4r_{\textrm{intra}}}$&$\frac{u_{00}^2}{2q^2v^2\tau^2+4r_{\textrm{inter}}}$\\
$C_{12,21}$&$-\frac{u_{00}^2}{2q^2v^2\tau_{00}^2}$&$-\frac{u_{00}^2}{2q^2v^2\tau^2+4r_{\textrm{intra}}}$&$-\frac{u_{00}^2}{2q^2v^2\tau^2+4r_{\textrm{inter}}}$\\
$C_{13,42}$&0&0&$\frac{u_{00}^2}{4q^2v^2\tau^2}$
\\
\hline\hline 
\end{tabular}
\label{tb_3}
\end{table}

For intervalley disorder potential, the resulting leading correction is calculated using Eq.~(\ref{eq_gaplessinter}) and Eq.~(\ref{eq_interlebtotbub}),
\begin{align}
\delta g_x^{\textrm{inter}}&\simeq e^2N_0v^2\tau^3\sum_{\bm{q}}\left(-16\frac{u_{00}^2}{4v^2q^2\tau^2}\right)\nonumber
\\
&\simeq-\frac{4e^2}{\pi}\int\frac{d^2q}{4\pi^2}\frac{1}{q^2}\nonumber
\\
&=-\frac{2e^2}{\pi^2}\int_{(v^2\tau\tau_{\phi})^{-\frac{1}{2}}}^{r_{\textrm{inter}}/v\tau}dq\frac{1}{q}\nonumber
\\
&=-\frac{e^2}{\pi^2}\ln\frac{r_{\textrm{inter}}^2\tau_{\phi}}{\tau}~,
\label{eq_dgxinter}
\end{align}
In this case, the negative logarithmic singularity appears. This reflects the fact that the intervalley scattering activates the \textit{effective} TRS relating the two valleys, leading to a WL correction.

\subsection{The LEB without inversion symmetry $(\lambda_R\neq0)$}

For comparison, we also consider a representative local two-valley Hamiltonian for the case of $\lambda_R\neq0$ and $\lambda_m\neq0$,
\begin{align}
H_{\textrm{LEB}}'=
\begin{pmatrix}
v_1\bm{k}\cdot\bm{\rho}&0\\
0&v_2(\bm{k}\cdot\bm{\rho})^{\intercal}
\end{pmatrix}~,
\end{align}
where neither \textit{physical} nor \textit{effective} TRS is present. Here $\bm{k}$ is measured relative to each Dirac node separately, and $v_1\neq v_2$ reflects the fact that no symmetry relates the two nodes. Because neither \textit{physical} nor \textit{effective} TRS constrains the disorder in this case, more scattering channels are symmetry-allowed than in the $T_{\textrm{LEB}}^*$-preserving case. The one constraint that remains is $(C_{2z}T)_{\textrm{LEB}}$, which must be preserved for the Euler bands to be well defined. Imposing it removes $\Sigma_z\Lambda_0$ and $\Sigma_z\Lambda_z$ from the intravalley set and $\Sigma_x\lambda_x$, $\Sigma_y\lambda_x$, $\Sigma_z\lambda_y$ and $\Sigma_0\lambda_x$ from the intevalley set, leaving the five intravalley and four intervalley disorder potentials that enter the scattering process. Furthermore, now two valleys have different velocity $v_1$ and $v_2$, scattering rates are also different. The two valleys have different velocities, so their densities of states at the Fermi level and the scattering rates differ as well. We define $r_F\equiv v_2/v_1$, and $N_{\alpha,0}$ $(\alpha=1,2)$ be the density of states at the Fermi level of the valley $\alpha$ (valley that is described by $v_\alpha$). Since $N_{\alpha,0}\propto v_{\alpha}^{-2}$, the two densities of states are related by $N_{2,0}=N_{1,0}/r_F^2$. The scattering rates of the two valleys therefore obey
\begin{align}
&\tau_{1,ij}^{-1}\equiv\pi N_{1,0}u_{ij}^2\equiv\tau_{ij}'^{-1},\nonumber
\\
&\tau_{2,ij}^{-1}=\frac{1}{r_F^2}\tau_{1,ij}^{-1}=\frac{1}{r_F^2}\tau_{ij}'^{-1}.
\end{align}
Using these, we define intravalley and intervalley scattering rate $\tau_{\textrm{intra}}'^{-1}$ and $\tau_{\textrm{inter}}'^{-1}$,
\begin{align}
&\tau_{\textrm{intra}}'^{-1}\equiv\tau_{xz}'^{-1}+\tau_{x0}'^{-1}+\tau_{yz}'^{-1}+\tau_{y0}'^{-1}+\tau_{0z}'^{-1}~,\nonumber
\\
&\tau_{\textrm{inter}}'^{-1}\equiv\tau_{xy}'^{-1}+\tau_{yy}'^{-1}+\tau_{zx}'^{-1}+\tau_{0y}'^{-1}~.
\end{align}
Throughout, $\tau_{\textrm{intra}}'^{-1}$ and $\tau_{\textrm{inter}}'^{-1}$ are expressed in terms of the scattering rates of the valley 1. We then calculate the conductivity correction due to disorder with $H_{\textrm{LEB}}'$ and these scattering rates, which is written as
\begin{align}
\delta g_x\simeq
\begin{cases}
\frac{2e^2}{\pi^2}\ln\frac{\tau_{\phi}}{\tau_{00}}~,&\textrm{Scalar disorder potential}~,
\\\\
\frac{2e^2}{\pi^2}\ln\frac{8r_{\textrm{intra}}'+5}{8r_{\textrm{intra}}'+5\tau'/\tau_{\phi}}~,&\textrm{Scalar+Intravalley disorder potential}~,
\\\\
\frac{2e^2}{\pi^2}\ln\frac{r_{\textrm{inter}}'+1}{r_{\textrm{inter}}'+\tau'/\tau_{\phi}}~,&\textrm{Scalar+Intervalley disorder potential}~,
\end{cases}
\label{eq_noTRSdgx}
\end{align}
where $r_{\textrm{intra}}'\equiv\tau'/\tau_{\textrm{intra}}'$ and $r_{\textrm{inter}}'\equiv\tau'/\tau_{\textrm{inter}}'$ with $\tau'^{-1}$ defined by $\tau'^{-1}\equiv\tau_{00}'^{-1}+\tau_{\textrm{intra}}'^{-1}$ and by $\tau'^{-1}\equiv\tau_{00}'^{-1}+\tau_{\textrm{inter}}'^{-1}$ for the scalar+intravalley  and scalar+intervalley disorder potential, respectively. Equation~(\ref{eq_noTRSdgx}) shows that scalar disorder potential still gives a WAL correction, whereas intravalley and intervalley disorder potentials remove the singular logarithmic contribution. This behavior is again consistent with the unitary class classification in Sec.~\ref{supsec_symAnal}.

\section{Comparison of the LEB and graphene}

The results of Sec.~\ref{supsec_dg} show that, for the $T^*$-preserving LEB, the localization correction coincides with that of graphene despite the different relative vorticities of the two Dirac nodes. We now make this correspondence explicit at the level of the Cooperon correction diagram. For graphene, we adopt the basis
\begin{align}
\label{eq_abbreviationGraph}
\{1,2,3,4\}=\{KA,KB,K'B,K'A\}~.
\end{align}
The only differences between graphene and the LEB lie in the forms of the low-energy Hamiltonians, Eqs.~(\ref{eq_swapGraph}) and (\ref{eq_swapLEB}), and in the antiunitary symmetries relating the two valleys. The
disorder matrices are otherwise taken in the same form. To make the comparison explicit, we first examine the weight factors $X_{\alpha\beta'}Y_{\beta\alpha'}$ appearing in Eq.~(\ref{eq_fullbubb}). Since these quantities are constructed from the disorder-averaged Green's function and the renormalized velocity vertices, they transform under the unitary matrix $U_{\textrm{exch}}=\mathbbm{1}_2\oplus\rho_x$ as
\begin{equation}
\label{eq_coeftrf}
U_{\textrm{exch}}X^{\textrm{LEB}}U_{\textrm{exch}}^{\dag}=X^{\textrm{graphene}}~,~~U_{\textrm{exch}}Y^{\textrm{LEB}}U_{\textrm{exch}}^{\dag}=Y^{\textrm{graphene}}~.
\end{equation}
We denote by $\bar{i}$ the index obtained under this exchange, namely
\begin{align}
\bar{1}=1~,\quad\bar{2}=2~,\quad\bar{3}=4~,\quad\bar{4}=3~.
\end{align}
Thus, the bar leaves the $K$ valley indices unchanged and exchanges the two pseudospin labels in the $K'$ valley block.

Since $X$ and $Y$ for both LEB and graphene are valley diagonal, each nonzero weight factor falls into one of three cases: both factors in the $K$ block, one factor in each block, or both factors in the $K'$ block. When all indices are in the $K$ valley block, $\alpha',\beta',\alpha,\beta\in\{1,2\}$, the transformation does not modify the matrix elements. Therefore, we obtain
\begin{align}
\label{eq_Kblocktrf}
X_{\alpha\beta'}^{\textrm{LEB}}Y_{\beta\alpha'}^{\textrm{LEB}}=X_{\alpha\beta'}^{\textrm{graphene}}Y_{\beta\alpha'}^{\textrm{graphene}}~,\qquad\alpha',\beta',\alpha,\beta\in\{1,2\}~.
\end{align}
We next consider the mixed case in which one factor belongs to the $K'$ block and the other to the $K$ block. If $\alpha,\beta'\in\{3,4\}$ while $\beta,\alpha'\in\{1,2\}$, then the bar acts only on the indices of $X$, giving
\begin{equation}
\label{eq_offDblocktrf}
X_{\alpha\beta'}^{\textrm{LEB}}Y_{\beta\alpha'}^{\textrm{LEB}}=X_{\bar{\alpha}\bar{\beta}'}^{\textrm{graphene}}Y_{\beta\alpha'}^{\textrm{graphene}}~.
\end{equation}
Conversely, if $\alpha,\beta'\in\{1,2\}$ and $\beta,\alpha'\in\{3,4\}$, the bar acts only on the indices of $Y$, so that
\begin{equation}
\label{eq_offDblocktrf2}
X_{\alpha\beta'}^{\textrm{LEB}}Y_{\beta\alpha'}^{\textrm{LEB}}=X_{\alpha\beta'}^{\textrm{graphene}}Y_{\bar{\beta}\bar{\alpha}'}^{\textrm{graphene}}~.
\end{equation}
Finally, when both factors lie in the $K'$ valley block, all indices are exchanged:
\begin{equation}
\label{eq_Kpblocktrf}
X_{\alpha\beta'}^{\textrm{LEB}}Y_{\beta\alpha'}^{\textrm{LEB}}=X_{\bar{\alpha}\bar{\beta'}}^{\textrm{graphene}}Y_{\bar{\beta}\bar{\alpha'}}^{\textrm{graphene}}~,\qquad\alpha',\beta',\alpha,\beta\in\{3,4\}~.
\end{equation}
Thus, every weight factor entering Eq.~(\ref{eq_fullbubb}) is mapped onto the corresponding graphene expression by the exchange $3\leftrightarrow4$ in the second valley.

We next compare the Cooperons themselves. For the scalar and intravalley disorder potentials, Cooperons connecting different valleys vanish. In these cases, the two valleys can be treated independently and the LEB is immediately found to be equivalent to graphene. The nontrivial case is the system with intervalley disorders, for which the relevant antiunitary symmetry differs between the two systems. In graphene, the intervalley disorder channels compatible with \textit{physical} TRS $T_{\textrm{graphene}}=\Pi_x\otimes\rho_x K=\Sigma_y\Lambda_y K$ are $\Sigma_x\Lambda_y,\Sigma_y\Lambda_y,\Sigma_z\Lambda_x$. In the LEB, the corresponding $T_{\textrm{LEB}}^*$-even intervalley disorder channels are $\Sigma_0\Lambda_y,~\Sigma_y\Lambda_y,~\Sigma_z\Lambda_x$. Applying $U_{\textrm{exch}}$ to the LEB disorder potential matrices maps them onto the graphene disorder matrices up to signs. These signs do not affect the Bethe-Salpeter kernel, because it depends only on the products of the form $V_{\alpha'\alpha}V_{\beta'\beta}$. Consequently, we obtain
\begin{equation}
\label{eq_Coopcomp}
C^{\textrm{LEB}}_{\alpha'\beta',\alpha\beta}=C^{\textrm{graphene}}_{\bar{\alpha'}\bar{\beta'},\bar{\alpha}\bar{\beta}}~.
\end{equation}

Finally, Eq.~(\ref{eq_fullbubb}) itself is symmetric under the exchange $3\leftrightarrow4$. Therefore, once the pseudospin labels in the second valley are exchanged, the full Cooperon correction diagram of the LEB becomes identical to that of graphene. We thus conclude that, within the Born approximation, the disorder-induced quantum correction to the conductivity $\delta g$ is independent of the relative vorticity of the two Dirac nodes. Instead, the decisive ingredient is the \textit{physical} or \textit{effective} antiunitary symmetry that constrains the Cooperon.

\end{document}